\documentclass[trackchanges,twocolumn]{aastex631}

\usepackage{threeparttable}
\usepackage{longtable}

\newcommand{\Ms}{M_{\odot}}

\newcommand{\MBH}{M_{\mathrm{BH}}}
\newcommand{\Mbulge}{M_{\mathrm{bulge}}}

\newcommand{\Mstar}{M_{\mathrm{star}}}
\newcommand{\Mdyn}{M_{\mathrm{dyn}}}

\newcommand{\Lbol}{L_{\mathrm{bol}}}

\newcommand{\sersic}{S\'{e}rsic }
\newcommand{\reff}{R\mathrm{_e}}

\newcounter{num}
\newcommand{\Ntwo}{\mathrm{N}\mathrm{I}\hspace{-1.2pt}\mathrm{I}}
\newcommand{\Stwo}{\mathrm{S}\mathrm{I}\hspace{-1.2pt}\mathrm{I}}
\newcommand{\Othree}{\mathrm{O}\mathrm{I}\hspace{-1.2pt}\mathrm{I}\hspace{-1.2pt}\mathrm{I}}
\newcommand{\Heone}{\mathrm{He}\ \mathrm{I}}

\newcommand{\Ha}{\mathrm{H}\mathrm{\alpha}}

\newcommand{\Pb}{\mathrm{Pa}\mathrm{\beta}}

 \newcommand{\mrm}{\ \mathrm}
 \newcommand{\AAA}{\ \mathrm{\mathring{A}}}

\begin{document}

\title{The Evolutionary Pathway of Low-mass Supermassive Black Holes at Intermediate Redshift: Insights from the JADES Survey}
\author[0009-0003-7134-0539]{Atsushi Hoshi}
    \affiliation{Astronomical Institute, Tohoku University, 6-3 Aramaki, Aoba-ku, Sendai, Miyagi 980-8578, Japan} 
    \affiliation{Institute of Space and Astronautical Science, Japan Aerospace Exploration Agency, 3-1-1, Yoshinodai, Chuou-ku, Sagamihara, Kanagawa, 252-5210, Japan}

\author{Toru Yamada}
\affiliation{Institute of Space and Astronautical Science, Japan Aerospace Exploration Agency, 3-1-1, Yoshinodai, Chuou-ku, Sagamihara, Kanagawa, 252-5210, Japan}
    \affiliation{Astronomical Institute, Tohoku University, 6-3 Aramaki, Aoba-ku, Sendai, Miyagi 980-8578, Japan} 





\keywords{Active galactic nuclei (16) --- Supermassive black holes (1663) --- Black hole physics (159) --- Seyfert galaxies (1447)}

\begin{abstract}

Understanding the relationship between supermassive black holes (SMBHs) and their host galaxies at different redshifts is crucial for unraveling the processes of SMBH-galaxy co-evolution.
We present the properties of nine type 1 Active Galactic Nuclei (AGNs) at intermediate redshift ($2<z<4$) using the JWST Advanced Deep Extragalactic Survey (JADES).
All of them show the significant $\mathrm{H\alpha}$ broad line and the AGN contribution in spectral energy distribution. 
Our sample covers SMBH masses of $10^{6.1-8.2}\ M_\odot$ and stellar masses of $10^{9.3-11.0}\ M_\odot$, comparable to those of the AGNs observed in the local universe. 
In the low-mass SMBH regime ($<10^{8}\ M_\odot$), the BH-to-stellar mass ratios in our sample ($0.01-0.1\%$) differ from those of the AGNs at $z>4$ ($1-10\%$), suggesting that black holes and galaxies may trace different evolutionary pathways at intermediate and high redshift.
We also perform 2D image decomposition using GALFIT to constrain the bulge mass by evaluating the bulge contribution in the rest-frame near-infrared flux.
We identify the AGNs with low BH-to-bulge mass ratios compared to those observed in the nearby bulge-dominant galaxies.
This finding suggests the existence of a galaxy-first evolutionary path, in which bulge formation occurs before substantial gas is efficiently accreted onto the central engine.
\end{abstract}

\section{Introduction}
Active galactic nuclei (AGNs) provide a unique opportunity to study the underlying processes that drive the growth of the central supermassive black holes (SMBHs) and the interaction with their host galaxies. 
The correlation between SMBH mass ($\MBH$) and bulge mass ($\Mbulge$) is well established in the local universe, providing strong evidence for the co-evolution of SMBHs and their host galaxies \citep[e.g.,][]{magorrian1998demography, haring_black_2004, kormendy2013coevolution,merritt2001m,mcconnell2013revisiting}.
However, whether $\MBH$ and $\Mbulge$ show a strong correlation in the distant universe remains an open question.
For instance, some studies have shown that some AGN populations deviate from the local $\MBH-\Mbulge$ relation \citep[e.g.,][]{Graham2015bh-bulge,ding2020mass,Sturm2024Abh-bulge}. 
On the other hand, observations of less luminous quasars have shown that the $\MBH-\Mbulge$ relation was already established as early as $z\sim6$ \citep[e.g.,][]{izumi2019subaru,izumi2021subaru}. 

From a theoretical perspective, different bulge formation mechanisms may influence the $\MBH-\Mbulge$ relation across cosmic time. 
Classical bulges are thought to have formed primarily through major mergers or \textit{in situ} starbursts \citep[e.g.,][]{Scannapieco2009model_disks,Clauwens2018insitu}.
Gas-rich mergers, for instance, are known to trigger both intense star formation and rapid SMBH growth, potentially leading to the formation of classical bulges \citep[e.g.,][]{Sanders1988ULIRG, hopkins2008cosmological,Brooks2016_bulge_merger}.
It has also been suggested that the coalescence of clumps in disk and the migration of disk stars to the central region contribute to the formation of classical bulges \citep[e.g.,][]{Noguchi1998Naturclump,Elmegreen2008clump,Park2019diskmigration}.
In contrast, pseudo-bulges which show a weaker correlation with SMBH mass may have evolved through secular evolution such as a disk instability \citep[e.g.,][]{Courteau1996secular,Kormendy2004Kennicutt,Orban2011NLRseyfert1_secular}. 
These systems grow primarily through the internal redistribution of disk material, leading to a more gradual evolutionary pathway.
Some cosmological simulations suggest that the progenitors of the pseudo-bulges form discs by the rapid supply of low angular momentum gas at $z>2$ \citep[e.g.,][]{Okamoto2013disk}.
Semi-analytic models suggest that overmassive BH systems at intermediate/high redshift compared to the local $\MBH-\Mbulge$ relation are primarily driven by galaxy mergers, which efficiently fuel SMBH growth \citep[e.g.,][]{croton2006evolution,Shirakata2019MNRAS,shimizu2024}. 
On the other hand, undermassive BH systems may be formed massive bulges through starbursts and migration of disk stars triggered by disk instabilities rather than mergers \citep{shimizu2024}.

Understanding the origin of SMBH–galaxy co-evolution requires studying AGNs hosting low-mass SMBHs ($\MBH < 10^8\ \Ms$) \citep[e.g.,][]{greene2012low}.
However, the available selection of such AGNs is limited due to their faintness.
Deep X-ray surveys and variability-based techniques are often used to detect low-luminosity AGNs \citep[e.g.,][]{Mountrichas2023wide, kimura_properties_2020, suh2020no, Hoshi,tanaka2025x}.
Recently, JWST has been demonstrated to be an effective means of identifying new AGNs that are difficult to detect with traditional methods.
For instance, new faint AGN candidates (also known as Little Red Dots: LRDs) that cannot be detected by X-rays or variability have been discovered through JWST observations \citep[e.g.,][]{Matthee2024lrd,Ananna2024lrd_xfaint,Kokubo2024Harikane,Labbe2025lrd}.
In addition, JWST/NIRSpec deep spectroscopic observations have revealed that high-z AGNs with low-mass SMBHs typically show overmassive BH systems compared to the local $\MBH - \Mstar$ relation \citep[e.g.,][]{harikane2023jwst, maiolino2023jades}. 
While the detection of the overmassive BH systems with low-mass SMBHs at high-z suggests the possibility of a rapid growth path of massive SMBHs, it remains unclear whether these systems are preferentially detected due to selection bias \citep[e.g.,][]{Pacucci_2023_violate, Li_2024_iceberg}.
Furthermore, since the low-mass SMBH sample is small even at intermediate redshift ($z < 4$), it is challenging to investigate the evolutionary pathways from the early universe to the final products observed in the local universe.

In this study, we utilize deep observational data from the JWST to investigate a rare sample of low-mass SMBHs at intermediate redshift. Through 2D image decomposition, we constrain bulge masses for our sample, aiming to explore the SMBH and bulge mass relation.

The structure of the paper is as follows. In section \ref{analysis}, we explain our sample and the analysis. In sections \ref{results} and \ref{discussion}, we present the results and discussion on the intermediate redshift AGNs from the JADES survey.   
Throughout this paper, we use a $\mathrm{\Lambda CDM}$ cosmological parameters of $\Omega_m=0.315$ and Hubble constant $H_0=67.4\mrm{km\ s^{-1} Mpc^{-1}}$ \citep{planck2020}. We use the AB magnitude system \citep{oke}.
\section{Analysis}\label{analysis}
\begin{table*}[t]
 \centering
 \caption{Basic Physical parameters of JWST intermediate redshift AGNs}
  \label{basic Physical parameters}
   \begin{tabular}{cccccccc}
    \hline \hline
NIRSpec ID &                   Tier &  NIRCam ID &  R.A. &  Decl. &  Redshift &      $f_\mathrm{AGN}$ &            X-ray  \\
      &  & & [deg] & [deg] & &   & \\
    \hline
    (1) & (2) & (3) & (4) & (5) & (6) & (7) & (8)\\
\hline
      22456 &      goods-n-mediumhst &    1026913 &  189.035785 &    62.243181 &  3.129330 & $0.07 \pm 0.17$ &           0\\
      23682 &      goods-n-mediumhst &    1093774 &  189.018339 &    62.250337 &  3.125450 & $0.19 \pm 0.25$ &            1 \\
      28074 &      goods-n-mediumhst &    1030801 &  189.064582 &    62.273823 &  2.261220 & $0.29 \pm 0.06$ &           1 \\
      29648 &     goods-n-mediumjwst &    1029648 &  189.209200 &    62.264259 &  2.959501 & $0.13 \pm 0.19$ &            1 \\
      78109 &     goods-n-mediumjwst &    1078109 &  189.279239 &    62.206856 &  2.920961 & $0.22 \pm 0.22$ &            1 \\
      80538 &      goods-n-mediumhst &    1027287 &  189.194736 &    62.246084 &  2.003074 & $0.78 \pm 0.23$ &           0 \\
      \hline
      49729 & goods-s-mediumjwst1180 &     206907 &   53.178496 &   -27.784104 &  3.189182 & $0.52 \pm 0.23$ &            0 \\
      51236 & goods-s-mediumjwst1180 &     208000 &   53.146160 &   -27.779940 &  2.583065 & $0.51 \pm 0.36$ &            0 \\
     209777 & goods-s-mediumjwst1180 &     209777 &   53.158471 &   -27.774046 &  3.708683 & $0.90 \pm 0.01$ &            0 \\
\hline
\end{tabular}
\begin{tablenotes}
\item[](1) NIRSpec ID (2) Tier (3) NIRCam ID (4)(5) Right Ascension and Declination determined by NIRCam (6) Spectroscopic redshift (7) Fraction of AGN luminosity to total infrared luminosity estimated from SED fitting (8) 0: X-ray detected, 1: X-ray undetected.
\end{tablenotes}
\end{table*}
\subsection{Sample selection}\label{sub-target selection}
We selected type 1 AGN candidates from the third data release of JADES consisting of calibrated spectra for 4000 galaxies \citep{2024JADESdr3}. 
We utilized spectra obtained with the medium-resolution grating G235M/F170LP ($\lambda/\Delta\lambda \sim 1000$) in the NIRSpec multi-object spectroscopy (MOS) mode, using the micro-shutter assembly (MSA) \citep{Ferruit2022jwstmos}.
A detailed data reduction of the JADES survey is given in \citet{JADES_overview_Eisenstein, Bunker2024JADES}. 
By adopting initial criteria that require a well-constrained spectroscopic redshift flag A (indicating that the redshift is determined by at least one emission line in the medium-resolution grating, $R=500$–$1500$) and detection of the $\Ha$ emission line regardless of its velocity width, we obtained 394 objects in the GOODS-S field and 740 objects in the GOODS-N field.
If we further restrict the sample to the objects at $z < 4$, 264 objects in GOODS-S and 571 in GOODS-N were selected.
We then checked the presence of a broad $\Ha$ emission line in broad line region (BLR) and the AGN contribution in the infrared wavelength as the criteria for identifying "broad-line" AGNs.
To assess the $\Ha$ line profile, we performed spectral fitting and selected objects that show a broad $\Ha$ emission line.
The presence of broad Ha component is confirmed by the statistical significance in Bayesian information criterion (BIC) test \citep{BIC_Schwarz}.
During the fitting process, we simultaneously measured the width of the narrow [$\Ntwo$]$\lambda\lambda6549,\ 6583$, $\Ha$, and [$\Othree$]$\lambda5007$ emission lines to distinguish AGN broad components from possible broadenings due to starburst activity or galactic outflow.
The spectral fitting procedures in the $\Ha$ region are described in Section \ref{subsection SMBH mass estimate}.
As the final criterion for selecting AGNs, we also performed SED fitting to investigate whether there is any significant contribution of the AGN component.
We describe the setup of SED fit in Section \ref{subsection-sed fitting}. 
We performed SED fitting on all sources with detected broad $\Ha$ emission.
Among these, one source (ID 00002916) showed an AGN fraction consistent with zero and was therefore classified as a star-forming galaxy. Another source (ID 00197911) exhibited broad $\Ha$ emission likely due to an outflow rather than a BLR, as inferred from its $[\Othree]$ line width. These two sources were excluded from the final AGN sample used in this study.
Finally, 3 objects in the GOODS-S field and 6 objects in the GOODS-N field satisfied the criteria of showing sufficiently high signal-to-noise ratios (S/Ns $>10$) for the $\Ha$ emission line and an AGN fraction greater than zero at $2<z<4$.
We show the basic physical parameters of our sample in Table \ref{basic Physical parameters}.

\subsection{Sample description}
Here, we describe additional properties of the AGN in our sample.
NIRSpec IDs 22456, 80538 49729, 51236, and 209777 are detected by the deep X-ray observations by Chandra \citep{2003Alexander_2Ms,2017Luo_7Ms}.
These are all classified as AGNs since they satisfied at least one of the six criteria, including high X-ray luminosity, hard spectra, X-ray-to-optical or radio excess, spectroscopic AGN features, or X-ray-to-NIR excess \citep[e.g.,][]{Xue2011ApJS_4Ms, 2017Luo_7Ms}. 
Broad near-infrared emission lines such as $\Heone$ ($\lambda=10833\AAA$) and $\Pb$ are shown in the spectra of IDs 23682, 28074, 80538, 49729, and 209777.
IDs 49729 and 209777 are also classified as AGN by \citet{Lyu_2024ApJL} by their SED analysis using the imaging data with JWST Mid-Infrared Instrument (MIRI).

\subsection{SMBH mass estimate}\label{subsection SMBH mass estimate}
\begin{figure*}[htbp]
    \centering
    \begin{tabular}{ccc}
        \begin{minipage}[t]{0.3\hsize}
            \centering
            \includegraphics[keepaspectratio, scale=0.35]{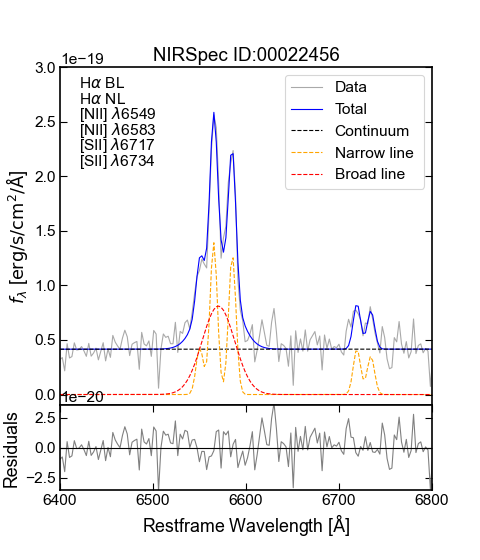}
        \end{minipage} &
        \begin{minipage}[t]{0.3\hsize}
            \centering
            \includegraphics[keepaspectratio, scale=0.35]{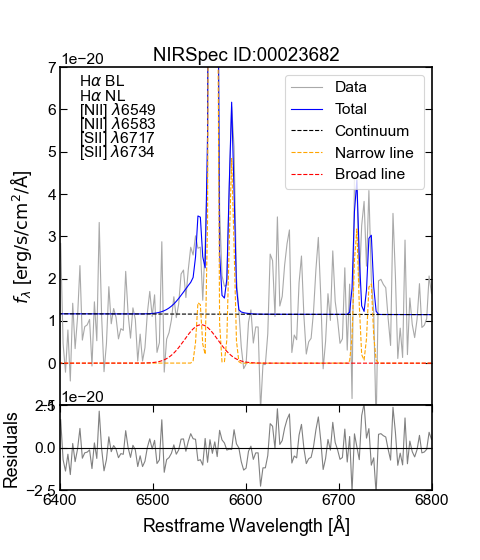}
        \end{minipage} &
        \begin{minipage}[t]{0.3\hsize}
            \centering
            \includegraphics[keepaspectratio, scale=0.35]{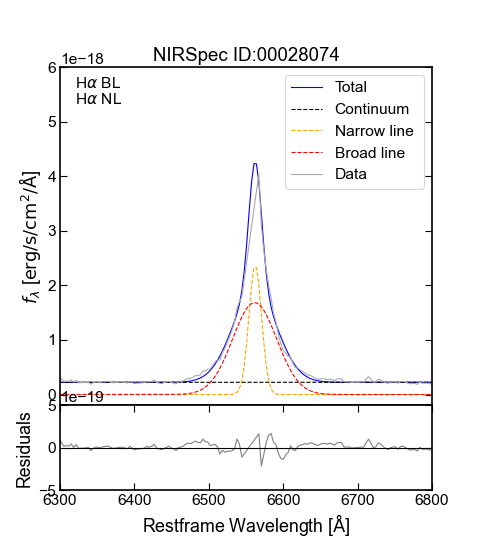}
        \end{minipage} \\
        \begin{minipage}[t]{0.3\hsize}
            \centering
            \includegraphics[keepaspectratio, scale=0.35]{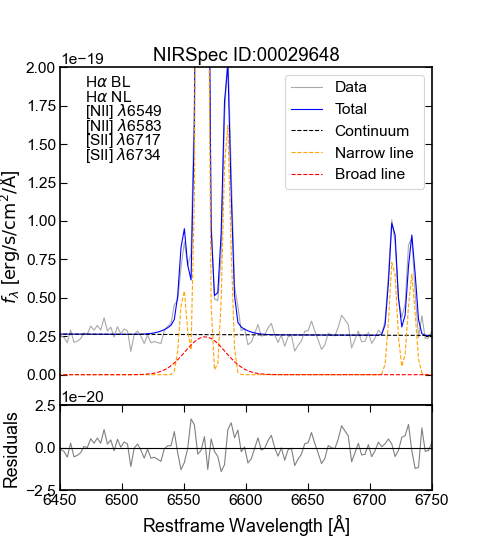}
        \end{minipage} &
        \begin{minipage}[t]{0.3\hsize}
            \centering
            \includegraphics[keepaspectratio, scale=0.35]{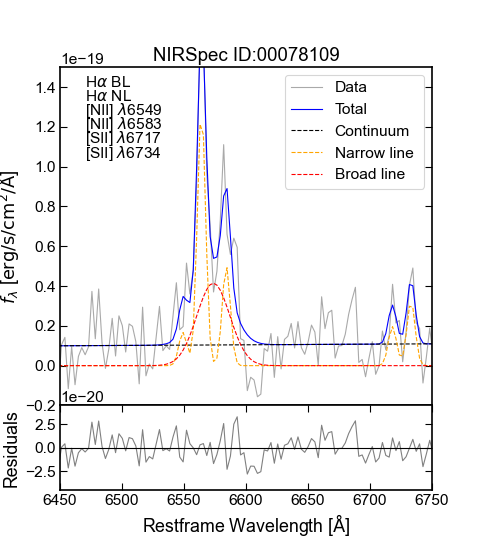}
        \end{minipage} &
        \begin{minipage}[t]{0.3\hsize}
            \centering
            \includegraphics[keepaspectratio, scale=0.35]{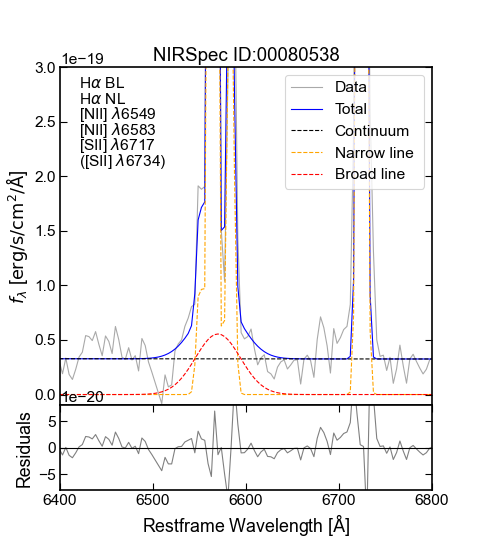}
        \end{minipage} \\
        \begin{minipage}[t]{0.3\hsize}
            \centering
            \includegraphics[keepaspectratio, scale=0.35]{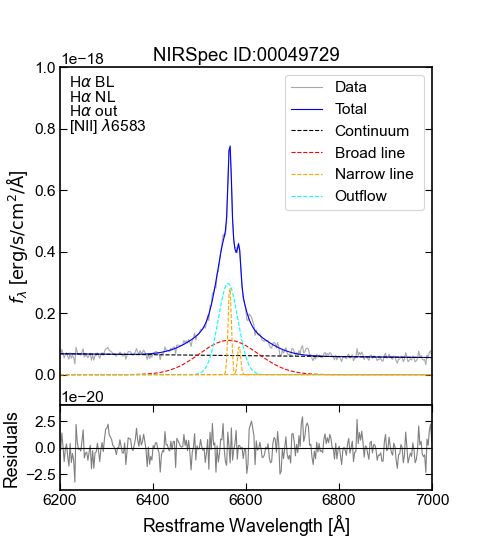}
        \end{minipage} &
        \begin{minipage}[t]{0.3\hsize}
            \centering
            \includegraphics[keepaspectratio, scale=0.35]{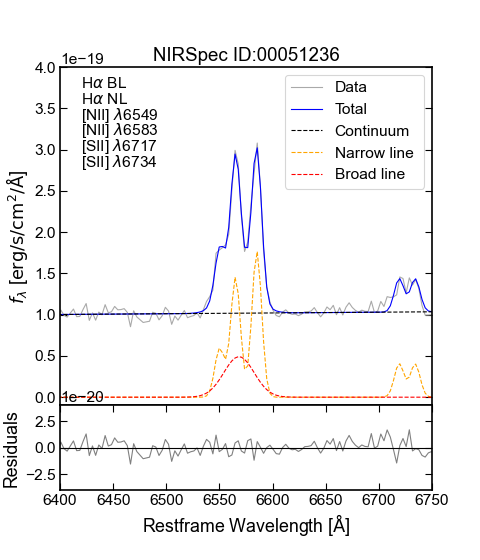}
        \end{minipage} &
        \begin{minipage}[t]{0.3\hsize}
            \centering
            \includegraphics[keepaspectratio, scale=0.35]{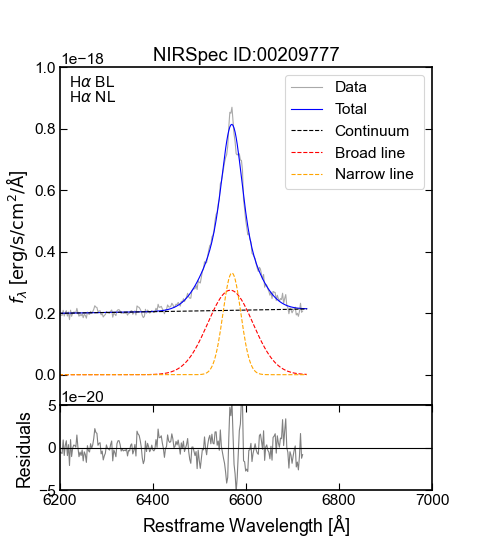}
        \end{minipage} \\
    \end{tabular}
    \caption{Spectral fitting around the $\Ha$ broad line for intermediate redshift AGNs. Gray and blue lines show the rest-frame observed spectra and the total fitting results respectively. Red, orange and black dashed lines represent the fitting components of the broad line, the narrow lines and the continuum respectively. Each NIRSpec ID is taken from \citet{2024JADESdr3}. The line components used in the fitting are labeled in the upper left corner of each panel. In the lower panel, we show the residual after subtraction of the total model from the data.}
    \label{fig-Spectral fitting}
\end{figure*}

Black hole masses are estimated based on the single-epoch virial method using the $\Ha$ broad line:\\
\begin{equation}
\label{mass_equation}
\begin{aligned}
&\MBH/\Ms =(2.0_{-0.3}^{+0.4}) \times10^6\\
&\quad
\times\left(\frac{L_{\Ha}}{10^{42}\ \mathrm{erg\ s^{-1}}}\right)^{0.55\pm0.02}\left(\frac{\Delta V_\mathrm{\Ha,BL}}{10^3\ \mathrm{km\ s^{-1}}}\right)^{2.06\pm 0.06}
\end{aligned}
\end{equation}
\\
where $L_{\Ha}$ is the total $\Ha$ line luminosity for the combined broad and narrow components, and $\Delta V_\mathrm{\Ha,BL}$ is the full width at half maximum for $\Ha$ broad emission lines \citep{greene2005estimating}.
For the object with ID 49729, we found the broad component associated with outflows, which is not included in the measurement of $L_{\Ha}$.
The H$\alpha$ profile of ID 00049729 requires two broad components, with velocity widths of $\sim 6700$ and $\sim 2200\ \mathrm{km\ s^{-1}}$. We interpret the broader component as originating from the gravitational potential of the central black hole and use it to estimate the black hole mass. The narrower broad component ($\sim 2200\ \mathrm{km\ s^{-1}}$) is likely associated with a galaxy-scale outflow and was therefore excluded from the $L_{\Ha}$ measurement used for the $L_{\Ha}$ calculation. Although the [$\Othree$] profile shows weak extended wings visually, the signal-to-noise ratio was insufficient to fit this outflow component, and H$\beta$ is not covered in the spectral range. While it is also possible that the narrower broad component originates from another SMBH, we interpret it as an outflow in this work, considering its profile is nearly symmetric and its peak is not shifted from narrow $\Ha$ line. The presence of two distinct broad components suggests different physical origins. If we use the narrower broad component for the black hole mass estimate for ID 49729, the resulting mass would be $ \MBH = 10^{7.1\pm0.1}\ \Ms$.
Here, we assess the uncertainties of black hole mass from different measurement methods. If we use only the broad $\Ha$ component in $L_{\Ha}$, black hole masses decrease by 0.13 dex on average.
There is also a difference in the estimated black hole mass when $L_{\Ha}$ in Equation \ref{mass_equation} is replaced with $L_{\lambda(5100)}$, the continuum luminosity at $5100 \AAA$ using Equation (1) from \citet{greene2005estimating}.
$L_{\lambda(5100)}$ can be converted from SED-based AGN bolometric luminosity and the bolometric correction for type 1 AGN of 10.33 \citep{richards2006spectral}. 
The BH masses of the SED-based AGN bolometric luminosity increase by 0.33 dex on average compared to those estimated from $L_{\Ha}$ for the combined broad and narrow components. 
For the measurement of $\Delta V_\mathrm{\Ha,BL}$, we perform a spectral fitting with the {\tt\string curve\_fit} tool from Python, a function from the {\tt\string scipy} module \citep{virtanen2020scipy}. We fit the continuum with a linear component and the emission lines with multi-Gaussian components in the $\Ha$ region of $6200 \AAA < \lambda_\mathrm{rest} < 6800 \AAA$, where $\lambda_\mathrm{rest}$ is the rest-frame wavelength.
The velocity widths of the narrow $\Ha$, [$\Ntwo$], and [$\Stwo$] lines are fixed. 
The [$\Ntwo$] 6585/6549 flux ratio is fixed to 3, based on the AGN sample of the Sloan Digital Sky Survey \citep{Ntwo3_2023}.
As the velocity width of the narrow [$\Othree$] line can differ from that of the narrow components in the $\Ha$ region, it is fitted independently.
In all objects, the $\Ha$ broad and narrow lines are separated well. 
The fitting results around the $\Ha$ line of our sample are shown in Figure \ref{fig-Spectral fitting}.

To obtain the intrinsic width of broad emission lines, we need to correct for the instrumental resolution.
The $\Ha$ emission line from objects at $z=2-4$ is covered by the spectroscopic data obtained with the medium resolution grating of G235M/F170LP with a nominal spectral resolution of 1000, corresponding to the typical instrumental broadening of 300 km s$^{-1}$.
We present the following results using the resolution-corrected broad $\Ha$  widths. 
The physical properties obtained in the fitting are listed in Table \ref{table Physical properties bh-st}. As the velocity widths of the narrow lines are comparable to the instrumental resolution, the values in Table \ref{table Physical properties bh-st} are reported without correction for instrumental broadening.

\begin{table*}[t]
 \centering
 \caption{Physical properties of JWST intermediate redshift AGNs}
  \label{table Physical properties bh-st}
  \begin{tabular}{cccccccccc}
    \hline \hline
  NIRSpec ID & $\Delta V_\mathrm{\Ha,BL}$ & $\Delta V_\mathrm{\Ha,NL}$ & $\Delta V_\mathrm{[\Othree],NL}$& $\log L_{\mathrm{H\alpha}}$  & $\log M_{\mathrm{BH}}$ &  SFR  & $\log M_{\star}$  & $\chi^2_\mathrm{red}$ \\
     &  [$10^3$km/s]&  [$10^3$km/s] &  [$10^3$km/s]&  [erg s$^{-1}$] &[$M_{\odot}$] &   [$M_{\odot}$/yr] &  [$M_{\odot}$] &   \\
     \hline
    (1) & (2) & (3) & (4) & (5) & (6) & (7) & (8) &(9)\\
   \hline
      22456 & $1.74 \pm 0.141$ & $0.43 \pm 0.036$  &           &          $41.65 \pm 0.04$ &                      $6.60 \pm 0.13$  &  $419.75 \pm 189.53$ &               $10.58 \pm 0.50$ &            $2.2(-99)$ \\
      23682 & $1.80 \pm 0.819$ & $0.27 \pm 0.005$ &   $0.31 \pm 0.011$&                     $41.55 \pm 0.02$ &                      $6.58 \pm 0.42$ &    $35.84 \pm 8.87$ &                $9.26 \pm 0.24$ &            $0.7(0.7)$ \\
      28074 & $3.10 \pm 0.049$ & $0.92 \pm 0.028$ &            &         $42.86 \pm 0.00$ &                      $7.79 \pm 0.11$ &      $0.03 \pm 0.11$ &               $10.84 \pm 0.02$ &            $4.1(8.1)$ \\
      29648 & $1.60 \pm 0.267$ & $0.31 \pm 0.003$ & $0.35 \pm 0.004$ &                     $41.85 \pm 0.01$ &                      $6.64 \pm 0.18$ &     $23.39 \pm 3.04$ &                $9.78 \pm 0.09$ &            $2.2(2.2)$ \\
     78109 & $1.25 \pm 0.262$ & $0.36 \pm 0.048$ &        &              $41.27 \pm 0.07$ &                      $6.10 \pm 0.22$ &     $41.10 \pm 27.61$ &                $9.99 \pm 0.36$ &            $4.2(4.5)$ \\
      80538 & $2.50 \pm 0.642$ & $0.36 \pm 0.004$ &          &           $41.83 \pm 0.02$ &                      $7.03 \pm 0.25$ &     $16.83 \pm 2.79$ &               $10.02 \pm 0.16$ &            $2.2(-99)$ \\
      \hline
      49729 & $6.79 \pm 0.293$ & $0.38 \pm 0.017$ & $0.46 \pm 0.006$&                     $42.24 \pm 0.02$ &                      $8.14 \pm 0.11$ &    $28.41 \pm 10.64$ &                $9.69 \pm 0.25$ &            $2.1(-99)$ \\ 
      51236 & $1.36 \pm 0.174$ & $0.49 \pm 0.020$ &        &             $41.29 \pm 0.05$ &                      $6.19 \pm 0.16$ &    $40.01 \pm 29.56$ &               $10.80 \pm 0.04$ &            $1.9(-99)$ \\
     209777 & $5.21 \pm 0.196$ & $2.06 \pm 0.079$ & $0.78 \pm 0.034$&                     $42.82 \pm 0.01$ &                      $8.23 \pm 0.11$ &      $0.03 \pm 2.24$ &               $11.02 \pm 0.02$ &            $3.3(-99)$ \\
\hline
\end{tabular}
\begin{tablenotes}
\item[]
(1) NIRSpec ID. 
(2) FWHM of the broad H$\alpha$ emission line.
(3) FWHM of the narrow H$\alpha$ line. 
No instrumental correction was applied.
(4) FWHM of the [$\Othree$] $\lambda5007$ narrow emission line, also uncorrected for instrumental broadening.
Blanks indicate either that the [$\Othree$] line is not covered by the medium-resolution grating or that the signal-to-noise ratio was insufficient for reliable measurement.
(5) Total H$\alpha$ luminosity, combining both broad and narrow components.
(6) Black hole mass 
(7) Star formation rate derived from SED fitting.
(8) Stellar mass from SED fitting.
(9) Reduced chi-square of the SED fit with AGN templates included. Values in parentheses denote the reduced chi-square when fitting without AGN contribution. A value of $-99$ indicates that the fit did not converge.
\end{tablenotes}
\end{table*}


\subsection{SED fitting}\label{subsection-sed fitting}
In this section, we describe the setup in the SED fitting code, CIGALE-v2022 \citep{boquien2019cigale, yang-xcigale}.
For the objects in GOODS-S field, we used UV to infrared photometry catalog \citep{Guo2013Goodss-cat} and X-ray point source catalogs for the 7 Ms exposure of the Chandra \citep{2017Luo_7Ms}.
The nominal detection limit of the hard X-ray band (2-7 keV) is $2.7 \times 10^{-17}\ \mathrm{erg\ cm^{-2}\ s^{-1}}$. 
For the objects in GOODS-N field, we used the CANDELS/SHARDS multi-wavelength photometry catalog \citep{Barro2019ApJS_goodsn_photo} and X-ray point-source catalogs for the 2 Ms exposure of the Chandra \citep{2003Alexander_2Ms,Xue2016ApJS2Ms250ks}.
The nominal detection limit of the hard X-ray band (2-8 keV) is $1.4 \times 10^{-16}\ \mathrm{erg\ cm^{-2}\ s^{-1}}$. 
Objects with X-ray detections are IDs 49729, 51236, 209777, 22456, and 80538. 
The actual detection limit for each individual source depends on their position within the Chandra field of view.
The photometric data observed with the JWST Near-Infrared Camera (NIRCam) in the filters, F090W, F115W,
F150W, F182M, F200W, F210M, F277W, F335M, F356W, F410M, F430M, F444W, F460M, F480M is taken from the archive catalog \citep{JADES_overview_Eisenstein, Eisenstein_JADES_dr2_photo, Rieke2023JADES}. 
Circular apertures with a radius of $0.35"$ (CIRC5) are used for all filters.
The circular aperture photometry is measured on PSF-convolved images, matched to the resolution of F444W.
To determine the optimal aperture size, we performed SED fitting using fluxes from all available aperture sizes (CIRC0–CIRC6).
We found that the CIRC5 aperture provided the best overall fitting results for our sample, yielding the smallest reduced chi-square values.
Here, we present the model module used for fitting in CIGALE.
For star formation history (SFH), we adopted the delayed star formation history model \citep{ciesla2017sfr}, which allows us to efficiently model early-type and late-type galaxies. And we adopted the stellar synthesis population of \citet{bruzual2003stellar} with a Salpeter initial mass function \citep{salpeter1955luminosity}.
We included the nebular emission from the photoionization models \citep{inoue2011rest}.   
We also used the dust attenuation model of \citet{Calzetti2000, Leitherer2002ApJS, Noll2009}, and the heated dust emission model of \citet{dale2014two}.
The SKIRTOR AGN model \citep{stalevski20123d,stalevski2016dust} was used for the AGN component. 
The best-fitting SED model is selected using the reduced chi-square value.
We also included SED models without an AGN component to assess whether the AGN fraction is necessary for the best fit.
For all objects, the inclusion of the AGN component leads to a best fit, supporting the presence of an AGN contribution. However, for the two objects (ID 23682 and ID 29648), the reduced chi-square values are nearly identical between the AGN and non-AGN models, indicating that the photometric SED fitting alone is insufficient to confirm the presence of an AGN.
Input models and parameters are summarized in Table \ref{SED CIGALE model}.
The SED fitting results are shown in Figure \ref{sed-plot1} and \ref{sed-plot2}.
For ID 80538, we excluded the data of IRAC channel 4 for SED fitting considering the excess of mid-infrared band from its nearby object.

\section{Results}\label{results}
\subsection{BH mass and Bolometric luminosity}
\begin{figure}[htbp]
\begin{center}
\includegraphics[width=90mm]{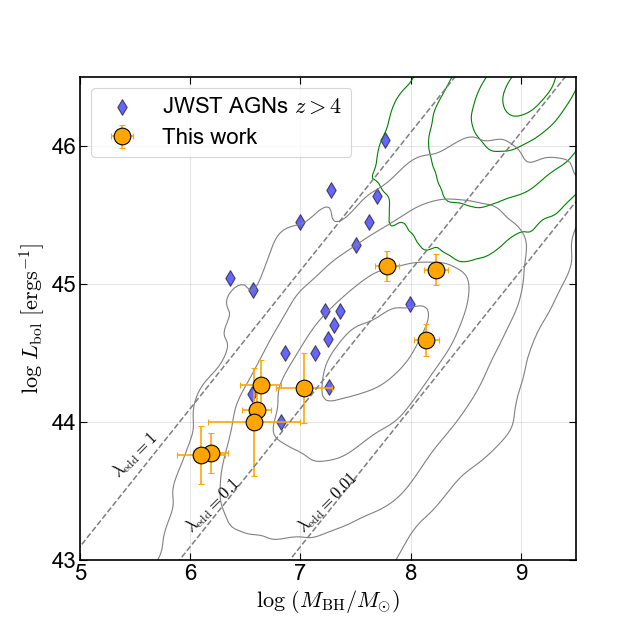}
\caption{$\MBH-\Lbol$ relation for our AGN sample. Gray and green contours represent the relation at $z<0.35$ \citep{liu2019comprehensive} and $1\leq z<2$ \citep{shen2011catalog} respectively. The blue diamonds represent AGN sample at $z>4$ \citep{harikane2023jwst,maiolino2023jades} from JWST. The dashed lines represent the Eddington ratios of $\lambda_\mathrm{edd} =1,\ 0.1,\ 0.01$.}
\label{lbol-bh}
\end{center}
\end{figure}

We show the relationship between BH mass and AGN bolometric luminosity in Figure \ref{lbol-bh}.
AGN bolometric luminosities are derived from the total $\Ha$ line luminosity for the combined broad and narrow components \citep{greene2005estimating}.
The gray and green contours represent the distribution of the objects in literature at $z<0.35$ \citep{liu2019comprehensive} and $1\leq z<2$ \citep{shen2011catalog}.
The bolometric luminosities are found to be $\Lbol=10^{44-45}$ erg s$^{-1}$ which is comparable to the low-z counterparts for the similar BH mass range \citep{liu2019comprehensive}. In contrast, our sample represents a fainter AGN population compared to those previously identified at $1\leq z<2$ \citep{shen2011catalog} and at $z>4$ \citep{harikane2023jwst, maiolino2023jades}.
The mean Eddington ratio of our intermediate redshift AGN sample is 0.17 which is consistent with that of typical X-ray selected type 1 AGNs \citep[e.g.,][]{suh2015edding}.

\subsection{Stellar mass and BH mass}
\begin{figure*}[htbp]
\begin{center}
\includegraphics[width=130mm]{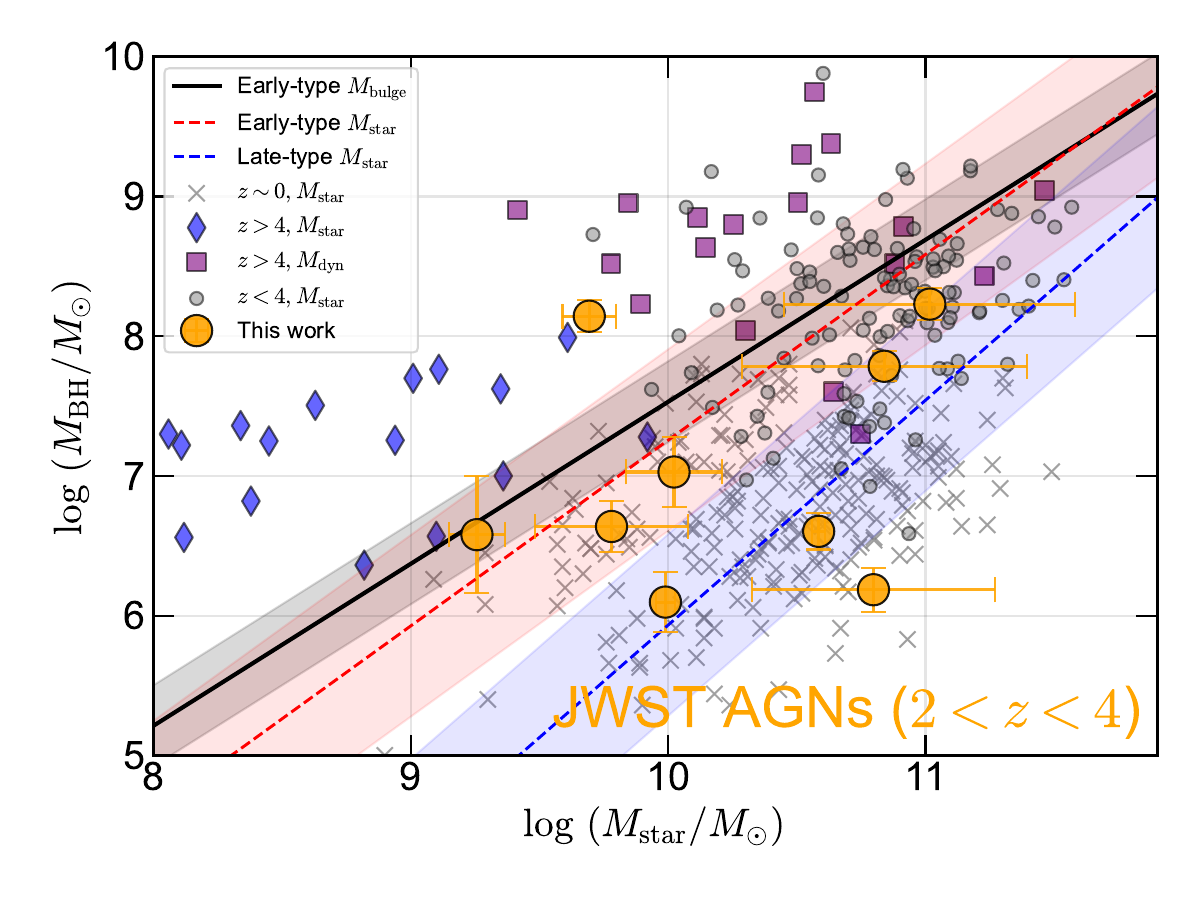}
\caption{$\MBH$ and $\Mstar$ in our AGN sample are shown by orange circles. The black solid line shows the $\MBH-\Mbulge$ relation for elliptical galaxies and bulge dominant galaxies in the local universe \citep{kormendy2013coevolution}. The red and blue dashed lines represent the $\MBH-\Mstar$ relation for early-type and late-type galaxies respectively \citep{Greene2020intermBH}. The gray, red, and blue shaded regions indicate the intrinsic scatter associated with each of these relations. The gray cross symbols represent the $\MBH-\Mstar$ relation for the AGN sample at $z<0.055$ \citep{reines2015relations}. The gray circles represent the $\MBH-\Mstar$ relation for variability-selected AGN sample at $z<4$ \citep{Hoshi}. The blue diamonds represent $\MBH-\Mstar$ for AGN sample at $z\sim4$ \citep{harikane2023jwst,maiolino2023jades} from JWST. The purple squares represent the $\MBH-\Mdyn$ relation for AGN sample at $z=4\sim7$ \citep{izumi2019subaru,pensabene2020alma,izumi2021subaru}.}
\label{bh-st}
\end{center}
\end{figure*}

Figure \ref{bh-st} shows the obtained $\MBH$ and $\Mstar$ of our intermediate-redshift AGN sample, represented by orange circles.
Our sample shows the similar $\MBH - \Mstar$ relation for low-z AGNs \citep{reines2015relations} except for ID 49729 with a high BH-stellar mass ratio of 0.03. 
Although the sample size is limited, the $\MBH$–$\Mstar$ relation in our AGN sample appears to show no strong redshift evolution, consistent with previous studies \citep[e.g.,][]{suh2020no,Hoshi,tanaka2025x}.
We highlight that most AGNs hosting low-mass SMBHs ($\MBH < 10^8\ \Ms$) at intermediate redshift have already formed enough stellar masses comparable to those of nearby galaxies or AGNs.
Furthermore, we found that there is a significant difference in BH-stellar mass ratios between high-z AGNs at $z>4$ ($1-10\%$) and our sample ($0.01-0.1\%$) in the low-mass SMBH regime.
This suggests that, in the low-mass SMBH regime, the undermassive BH systems (with massive hosts) at intermediate redshift may be in a different evolutionary pathway from overmassive BH systems at high redshift.

\subsection{BH-Galaxy Co-evolution}
\begin{figure}[t]
    \centering
    \includegraphics[width=0.35\textwidth, keepaspectratio]{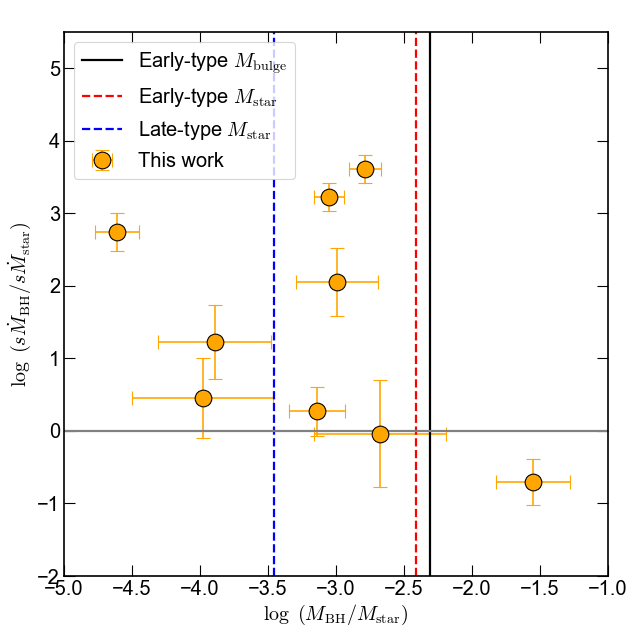}
    \caption{$M_\mathrm{BH}/M_\mathrm{star} -s\dot{M}_\mathrm{BH}/s\dot{M}_\mathrm{star}$ diagram. Black solid line represents the $\MBH/\Mbulge$ ratio for early-type galaxies with $\Mbulge=10^{11}\ \Ms$ in local universe \citep{kormendy2013coevolution}. Red and blue doted lines represent the $\MBH/\Mstar$ ratios for early-type and late-type galaxies with $\Mstar=10^{11}\ \Ms$ in local universe \citep{Greene2020intermBH}.}
    \label{fig-sacc-ssfr}
\end{figure}

In this section, we investigate the co-evolution between SMBHs and host galaxies in our sample.
The black hole growth rate is described as
\begin{equation}
\label{equation-accreion}
\dot{M}_\mathrm{BH} = \frac{(1-\eta)\Lbol}{\eta c^2},
\end{equation}
\\
where $\eta=0.1$ is the radiative efficiency in the case of black holes, and $c$ is the speed of light. $\Lbol$ is converted from the total $\Ha$ line luminosity. If the Eddington ratio is constant, SMBH mass grows exponentially from its initial mass.
Here, we consider the specific black hole growth rate, $s\dot{M}_\mathrm{BH}=\dot{M}_\mathrm{BH}/\MBH$ and the specific star formation rate, $s\dot{M}_\mathrm{star}=SFR/\Mstar$, where $SFR$ is star formation rate derived from the SED fitting as described in Section \ref{subsection-sed fitting}.
The ratio $s\dot{M}_\mathrm{BH}$ to $s\dot{M}_\mathrm{star}$ provides insight into the growth direction in $\MBH-\Mstar$ plane.
For example, if an object shows $(s\dot{M}_\mathrm{BH}/s\dot{M}_\mathrm{star}) > 1$, it indicates that SMBHs are growing faster than their host galaxies, potentially leading to the formation of overmassive BH systems.
Figure \ref{fig-sacc-ssfr} shows the $M_\mathrm{BH}/M_\mathrm{star} -s\dot{M}_\mathrm{BH}/s\dot{M}_\mathrm{star}$ diagram in our intermediate redshift AGN sample.
Black vertical line represents the $\MBH/\Mbulge$ ratio taken from \citet{kormendy2013coevolution}, red and blue dotted lines represent the $\MBH/\Mstar$ ratio in early-type and late-type galaxies with $\Mstar=10^{11}\ \Ms$ in the local universe \citep{Greene2020intermBH}.
We found that seven AGNs with accretion-dominated systems ($s\dot{M}_\mathrm{BH} > s\dot{M}_\mathrm{star}$), located above the gray horizontal line, tend to host undermassive SMBHs located to the left of the black line. These AGNs are growing their black holes more rapidly than their host galaxies at the observed epoch, and given their low $\MBH/\Mstar$ ratios, they are evolving toward the established BH–bulge relation observed in the local universe.
On the other hand, the AGN hosting overmassive SMBH at the right of the black line shows a star-forming dominated system. 
Given its high $\MBH/\Mbulge$ ratio, it is also evolving toward the local BH–bulge relation, as the host galaxy is growing faster than its central black hole.
The AGN with similar specific growth rate for the black hole and star formation ($s\dot{M}_\mathrm{BH} \sim s\dot{M}_\mathrm{star}$), located along the gray line, may trace a synchronized evolutionary path and evolve into the typical galaxies (elliptical, early-type, or late-type) observed in the local universe without significantly changing their $M_\mathrm{BH}/M_\mathrm{star}$ ratio.
The $M_\mathrm{BH}/M_\mathrm{star}$ ratios in our sample are regulated by the different growth rates of black holes and their host galaxies, evolving toward typical massive galaxies in the local universe.
\section{Discussion}\label{discussion}
\subsection{2D image decomposition}
\begin{figure*}[htbp]
    \centering
    \setlength{\tabcolsep}{0pt} 
    \renewcommand{\arraystretch}{0} 
    \begin{tabular}{l}
        \includegraphics[height=2.3cm, keepaspectratio]{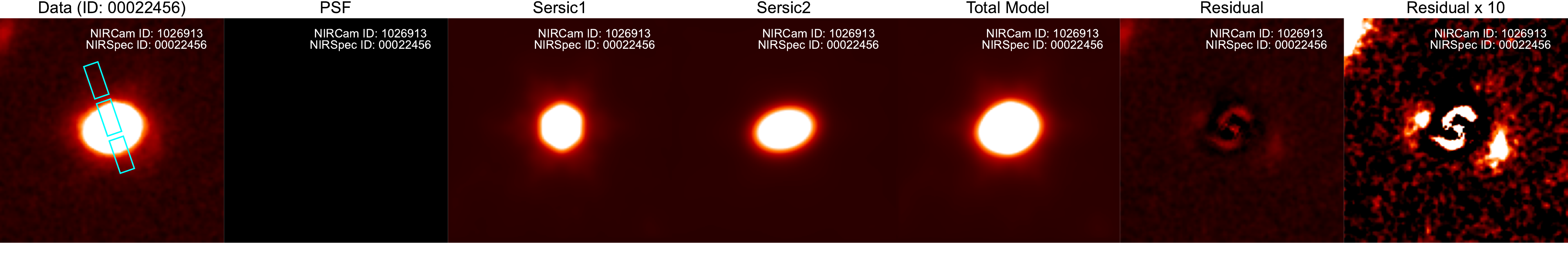} \\
        \includegraphics[height=2.3cm, keepaspectratio]{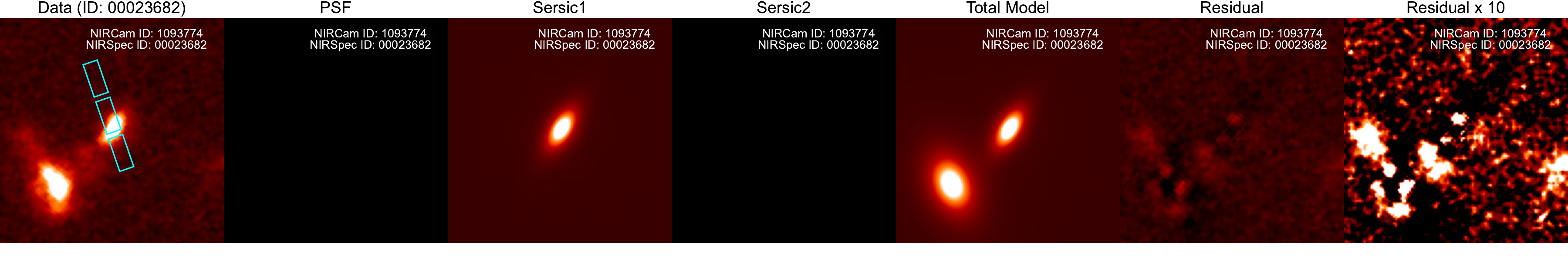} \\
        \includegraphics[height=2.3cm, keepaspectratio]{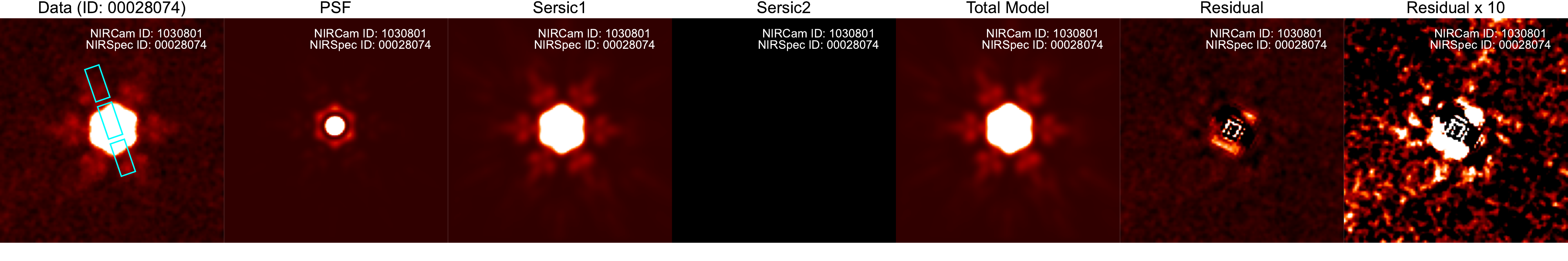} \\
        \includegraphics[height=2.3cm, keepaspectratio]{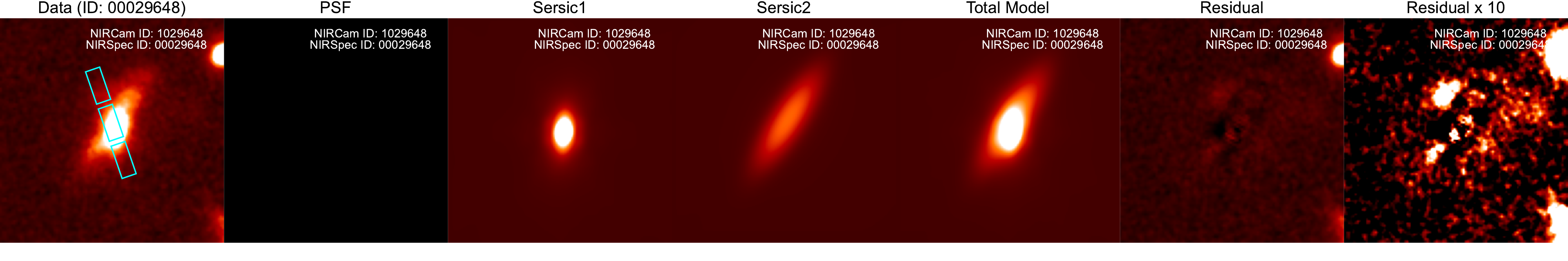} \\
        \includegraphics[height=2.3cm, keepaspectratio]{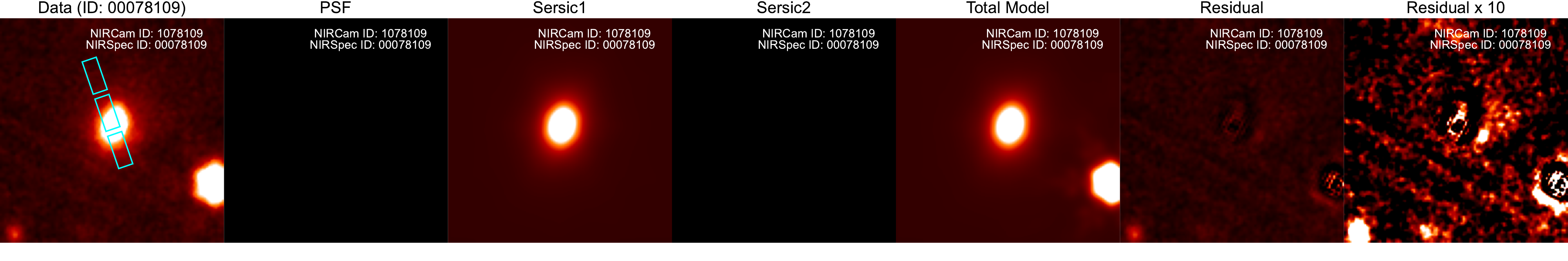} \\
        \includegraphics[height=2.3cm, keepaspectratio]{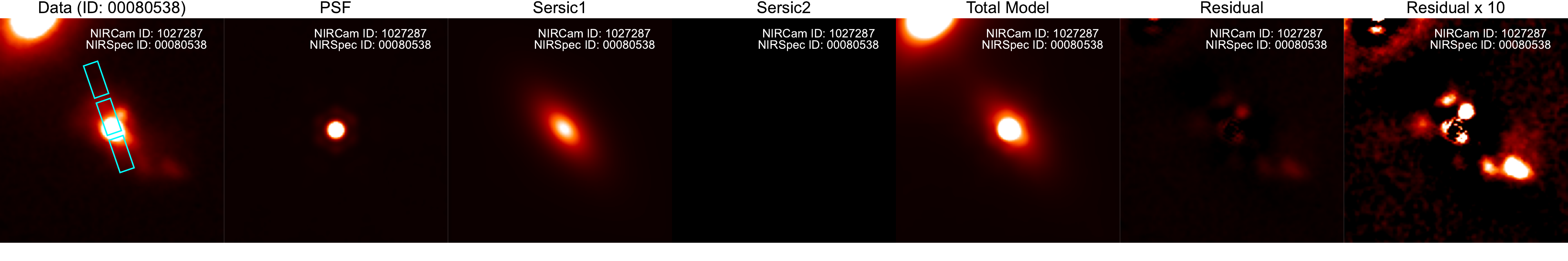} \\
        \includegraphics[height=2.3cm, keepaspectratio]{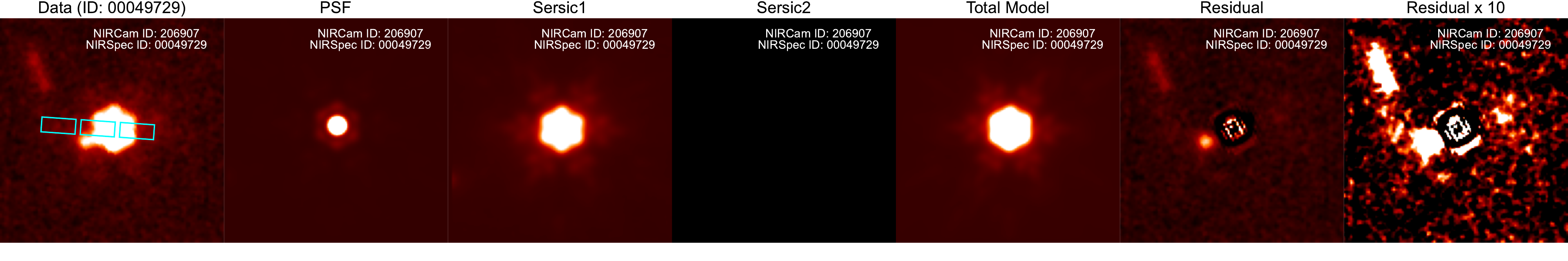} \\
        \includegraphics[height=2.3cm, keepaspectratio]{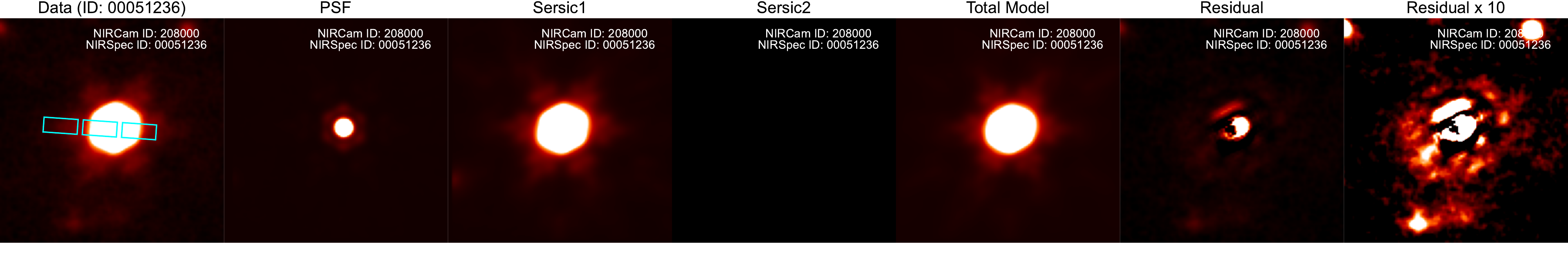} \\
        \includegraphics[height=2.3cm, keepaspectratio]{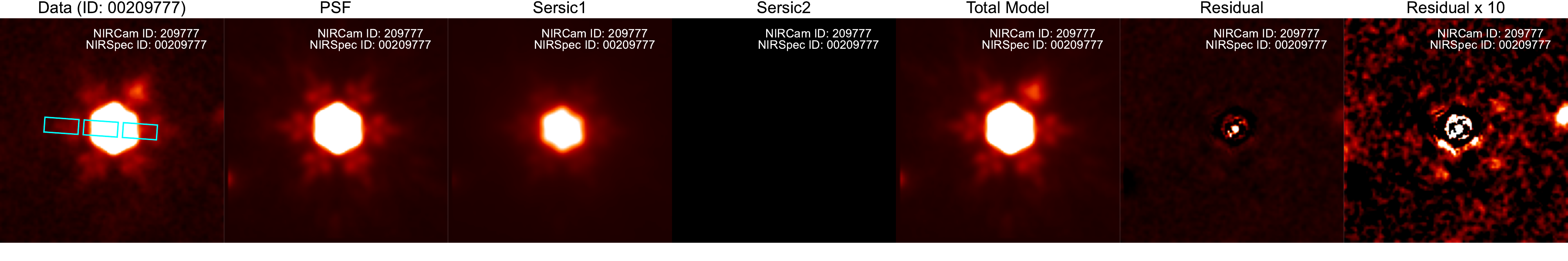} \\
    \end{tabular}
    \caption{The results of 2D image decomposition performed using Galfit. Images are 100$\times$100 pixel cutouts displayed with a logarithmic red color scale. The leftmost column shows imaging data in NIRCam/F444W. Cyan rectangles indicate the positions of the three MSA slits. Each slit has a size of $0.20'' \times 0.46''$. The second to fourth columns show the PSF, 1st \sersic, and 2nd \sersic components, respectively. The total model is shown in the fifth column, and the residuals between the data and total model are in the sixth column. The rightmost column represents the ten times residuals. Nearby galaxies within the cutout (IDs 23682, 78109, and 80538) were simultaneously modeled with single \sersic profiles to account for de-blending.}
    \label{Galfit fig components}
\end{figure*}

Using high-quality imaging data from JWST, we performed 2D image decomposition on our sample to evaluate the bulge contribution to the total stellar mass.
2D image decomposition analysis can be performed by either parametric or non-parametric methods. However, for AGNs with a prominent PSF structure (particularly in JWST data), or for galaxies containing nearby objects or clumps, non-parametric methods are considered to be less effective. Therefore, in this study, the host galaxies and AGN contribution are analyzed using a parametric approach by adopting the \sersic profile \citep{Sersic1968} and theoretical PSF model generated by the Python package {\tt\string WebbPSF} \citep{perrin2012webbpsf,Perrin2014webbpsf}.
We conducted 2D image decomposition using Galfit \citep{Peng_2002AJ_galfit,peng2010detailed}. 
We used 100$\times$100 pixel cutouts of the NIRCam drizzled images taken in the filters, F115W, F150W, F277W, and F444W.
We tested the following fitting models: single PSF, single \sersic, PSF $+$ \sersic, and double \sersic components. The best-fit model was determined based on the BIC test. 
We obtained the best results in the rest-frame NIR filter of F444W, where the bulge contribution is dominant.
To ensure consistency across the sample, we adopted F444W for all sources, which also provides stable PSF characterization and minimizes resolution-dependent systematics in the decomposition. For two lower-redshift AGNs (ID 28074 and ID 49729 at $z=2.0,\ 2.3$),  we performed 2D decomposition using F277W images.
The fitting results obtained from F277W are in good agreement with those from F444W.
The result of the 2D image decomposition analysis with Galfit is summarized in Table \ref{table galfit AGNfrac}. 
The individual fitting components generated by Galfit are shown in Figure \ref{Galfit fig components} and \ref{fig-SB}. 

IDs 28074, 49729, and 209777 show PSF-like features in their \sersic models, since their effective radii are comparable to the FWHM of the PSF (2.22 pixels), corresponding to 0.14 arcsec or $\sim 1$ kpc at $z=2-4$, which is a physically plausible size for a compact host galaxy component.
For IDs 23682, 78109, and 80538, we also fitted the surrounding galaxies or clumps by the \sersic models simultaneously to account for de-blending.
The PSF component was not significantly detected in IDs 22456, 23682, 29648, and 78109.
Our analysis reveals that the systems in our sample are stellar-dominated in the rest-frame NIR.
For IDs 22456 and 29648, the best fit was obtained with double \sersic components, allowing us to decompose the galaxy profiles into bulge and disk components.
Among the objects whose host galaxy component is best fitted by a single \sersic profile, IDs 49729 and 209777 have the indices typically classified as early-type galaxies ($n > 2$) indicating that they are likely to be bulge-dominated systems.
In contrast, host galaxies of IDs 23682, 28074, 78109, and 51236, were typically classified as late-type galaxies ($n < 2$).
We also checked the consistency between the 2D image decomposition analysis and SED fitting using the AGN fraction of the total flux at F444W.
The AGN fraction from SED fitting ($f_\mathrm{SED}$) is estimated from the ratio of AGN emission to the total model.
The AGN fraction from 2D image decomposition ($f_\mathrm{2D}$) is calculated from the PSF fraction of the total model.
The $f_\mathrm{SED}$ and $f_\mathrm{2D}$ coincide within 0.1, except for ID 78109.
For the discrepancy in the result for ID 78109, we interpret the main reason to be the uncertainty in SED fitting, caused by the low S/N data in rest-UV to optical range (or the time-dependent photometry due to AGN variability).

\subsection{Constraint for bulge mass}
\begin{figure}[t]
    \centering
    \includegraphics[width=0.4\textwidth, keepaspectratio]{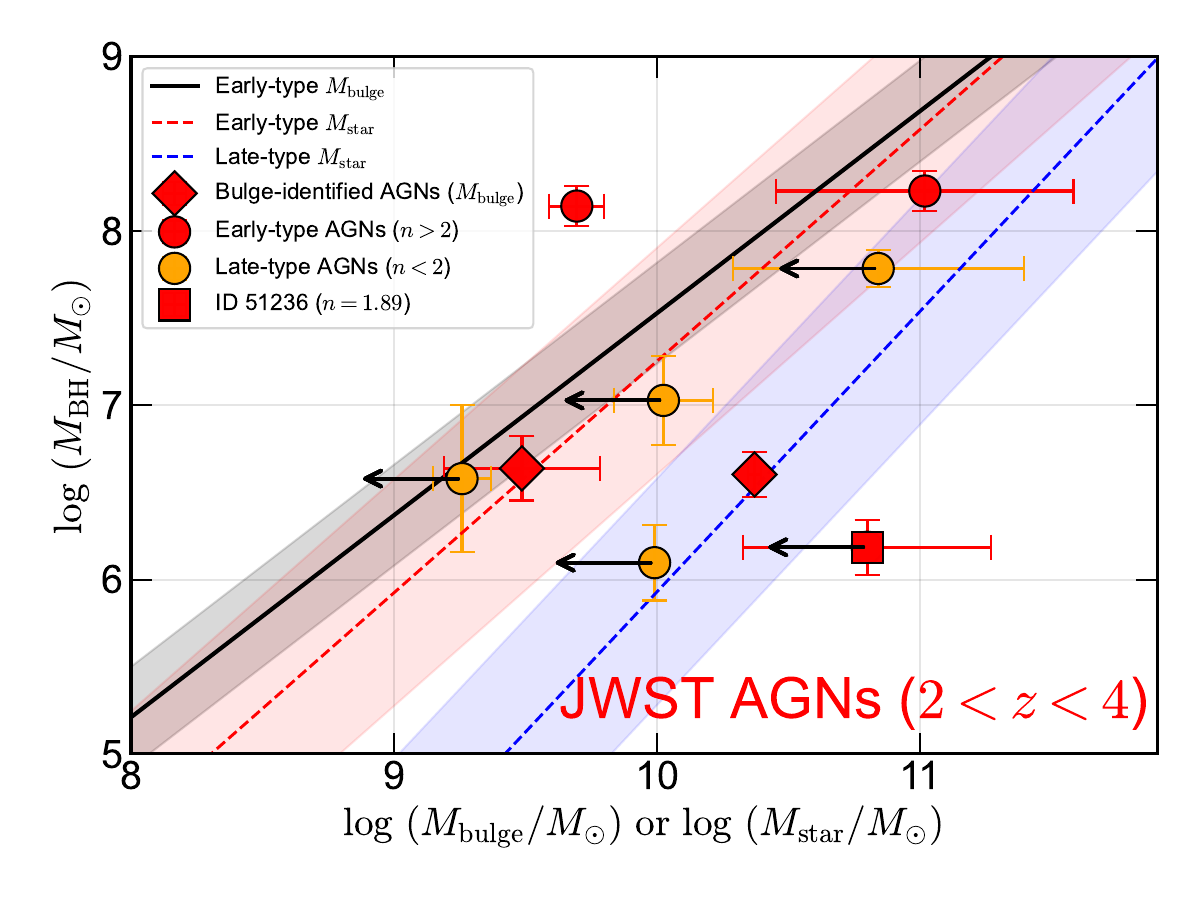}
    \caption{Black hole mass ($\MBH$) vs host galaxy bulge mass ($\Mbulge$). Red diamonds represent the $\Mbulge$ estimated AGNs. Red and orange circles represent the early-type and late-type AGNs. ID 51236, which has a bulge component but an unconstrained $\Mbulge$, is represented by a red square. Black allows show the upper limit of bulge mass. The black solid line shows the $\MBH-\Mbulge$ relation for elliptical galaxies and bulge dominant galaxies in the local universe \citep{kormendy2013coevolution}. The red and blue dashed lines represent the $\MBH-\Mstar$ relation for early-type and late-type galaxies respectively \citep{Greene2020intermBH}. The gray, red, and blue shaded regions indicate the intrinsic scatter associated with each of these relations.}
    \label{bh-bulge}
\end{figure}

Constraining bulge mass at intermediate redshifts provides unique insights into the co-evolution of black holes and galaxies, since diverse bulge formation mechanisms should have affected the $\MBH-\Mbulge$ relation.
In this section, we discuss the bulge mass for our sample inferred from the results of the 2D-fitting.
The bulge mass was measured by three methods:
(1) For the objects that can be decomposed to bulge and disk, the bulge mass was calculated from $\Mbulge\sim B/T\times\Mstar$, where $B/T$ is the bulge to total ratio converted from the bulge to total (bulge$+$ disk) luminosity ratio in rest-NIR band.
(2) For bulge-dominated systems ($n>2$), their bulge masses were approximated to the stellar masses ($\Mbulge\sim \Mstar$).
(3) For late-type AGNs ($n<2$), we treat the stellar mass as an upper limit of the bulge mass ($\Mbulge < \Mstar$). This is because, when a galaxy is fitted with a single \sersic profile with $n < 2$, it is difficult to isolate the bulge contribution especially if the galaxy has multiple structural components such as a bulge, disk, or clumps. 
IDs 78109 and 51236, however, are considered to possess bulge structures as their \sersic indices ($n=1.86,\ 1.89$) are hardly explained by a pure exponential disk profile ($n=1$).
For ID 51236, the consistency between the 2D imaging decomposition and the SED fitting supports the interpretation that a bulge may be present.
Figure \ref{bh-bulge} shows the resulting $\MBH-\Mbulge$ relation. AGNs with a bulge component are highlighted in red.
We found that most of our sample shows undermassive BH systems relative to the local $\MBH-\Mbulge$ relation.
The presence of these systems with small $\MBH/\Mbulge$ ratios at intermediate redshift suggests the existence of a bulge-first evolutionary pathway.

Most massive galaxies ($\Mstar \ge 10^{10}\ \Ms$) are thought to possess a prominent bulge \citep[e.g.,][]{Benton2024ApJ_bulge}.
Given this, most of the undermassive BHs at intermediate redshift including orange symbols (IDs 28074, 78109, and 80538) may also host a bulge component, as they have enough stellar masses despite showing small \sersic indices ($n<2$).
However, since their bulge masses cannot be constrained, it's difficult to directly compare them with bulge-dominated systems in the local universe.

\subsection{Bulge formation scenarios}
Multiple scenarios have been proposed for bulge formation, which are broadly categorized into two types: classical bulges and pseudo-bulges.
The overmassive BH systems (ID 49729), could have experienced the major merger which triggered both BH growth and classical bulge formation \citep[e.g.,][]{hopkins2008cosmological, croton2006evolution}.
Conversely, we also showed the existence of undermassive BH systems at intermediate redshift. 
These low $\MBH/\Mbulge$ ratios may be explained by disk instabilities, which play a significant role in the growth of both black holes and (pseudo-)bulges \citep[e.g.,][]{shimizu2024}.
As an alternative scenario, they could have evolved through the migration of disk stars or clumps to the bulge over long timescales \citep[e.g.,][]{Noguchi1998Naturclump,Elmegreen2008clump, Park2019diskmigration}.
However, forming a dense central structure ($n > 2$) characterized as a classical bulge, likely requires not only secular processes but also growth mechanisms that build bulges on short timescales without significantly accreting gas onto the black hole in the early epoch.
It is suggested that AGNs with small $\MBH/\Mbulge$ ratios at intermediate redshift have experienced a diverse bulge formation process.

\section{Conclusion}
We identified nine Type 1 AGN candidates at intermediate redshift ($2 < z < 4$) from the JADES survey based on the detection of broad H$\alpha$ emission lines and non-zero AGN contributions derived from SED fitting.
$\MBH$ is estimated using the single-epoch virial method applied to the broad H$\alpha$ line. The resulting $\MBH$–$L_{\mathrm{bol}}$ and $\MBH$–$\Mstar$ relations are broadly consistent with those observed in low-z AGN samples. We found a significant difference in the $\MBH$/$\Mstar$ ratios between our sample and high-redshift AGNs ($z > 4$) in the low-mass SMBH regime ($\MBH<10^8\ \Ms$).
We also explored the evolution path of black holes and their host galaxies by comparing the black hole growth rate to the specific star formation rate. The results for our sample showed that black holes and their hosts are evolving toward the population of massive galaxies in the local universe.
Using 2D image decomposition of JWST/NIRCam F444W data, we constrained the bulge contribution in the rest-frame near-infrared flux. We found the presence of AGNs with lower BH-bulge mass ratios than those of the nearby bulge-dominant galaxies, suggesting that some AGNs at intermediate redshift may represent a galaxy-first evolutionary stage.
Previous studies of the distant universe have suggested the potential for a rapid growth path of massive SMBHs.
In this study, we report the first discovery of an undermassive BH system with a prominent bulge at intermediate redshift, utilizing deep spectroscopic and imaging data from the JADES survey.
Our findings reveal the possibility of a galaxy-first evolutionary path, in addition to the commonly observed BH-first path.

\begin{acknowledgments}

This work is based on observations made with the NASA/ESA/CSA James Webb Space Telescope.
The data were obtained from the Mikulski Archive for Space Telescopes at the Space Telescope Science Institute, which is operated by the Association of Universities for Research in Astronomy, Inc., under NASA contract NAS 5-03127 for JWST. These observations are associated with program 1180, 1181, 1210, 1286, 3215.
The data is available at the Mikulski Archive for Space Telescopes: \dataset[doi:10.17909/8tdj-8n28]{{doi:10.17909/8tdj-8n28}} and \dataset[doi:10.17909/fsc4-dt61]{{doi:10.17909/fsc4-dt61}}.
This work was supported by JST, the establishment of university fellowships towards the creation of science technology innovation, Grant Number JPMJFS2102.
This work was also supported by JST SPRING, Grant Number JPMJSP2114.
This publication is based upon work supported by KAKENHI (22K03693) through Japan Society for the Promotion of Science.

\end{acknowledgments}

\bibliography{main}{}
\bibliographystyle{aasjournal}

\appendix

\begin{table*}[ht]
 \centering
 \caption{SED CIGALE model parameter}
  \label{SED CIGALE model}
   \begin{tabular}{ll}
    \hline \hline
   
  Parameter   & Input values \\
   \hline 
   \multicolumn{2}{c}{Star Formation History : sfhdelayed \citep{ciesla2017sfr} }\\
    e-folding time of the main stellar population model in Myr
    & 50, 100, 500, 1000, 2000, 5000, 10000, 12000 \\
   Age of the main stellar population in the galaxy in Myr
    & 50, 100, 500, 1000, 2000, 5000, 10000, 12000  \\
   e-folding time of the late starburst population model in Myr
    & 50  \\
   Age of the late burst in Myr. The precision is 1 Myr
    & 50  \\
   Mass fraction of the late burst population
    & 0.1  \\
    \hline 
    \multicolumn{2}{c}{Stellar Synthesis Population : bc03 \citep{bruzual2003stellar}}\\
    Initial Mass Function
    & Saplpeter \citep{salpeter1955luminosity}   \\
    Metallicity     & 0.02 \\
    \hline 
    \multicolumn{2}{c}{Nebular emission : nebular \citep{inoue2011rest}}\\
    Ionisation parameter     & $-0.2$ \\
    Line width in km/s &    300.0\\
    \hline
     \multicolumn{2}{c}{Dust attenuation : dustatt$\_$modified$\_$starburst \citep{Calzetti2000, Leitherer2002ApJS,Noll2009}}\\
    Color excess of the nebular lines : $E(B-V)_\mathrm{lines}$  & 0, 0.1, 0.3, 0.6, 1, 2, 3 \\
    Color excess of the stellar continuum : $E(B-V)_\mathrm{f}$  & 0.44 \\
    Slope delta of the power law modifying the attenuation curve & $0.0, -0.25, -0.5$ \\
    \hline
     \multicolumn{2}{c}{Dust emission : dale2014 \citep{dale2014two}}\\
    Powerlaw slope $dU/dM \propto U^\alpha$ & 2.0\\
    \hline
     \multicolumn{2}{c}{AGN model : skirtor2016 \citep{stalevski20123d,stalevski2016dust}}\\
    
    Inclination  & 0, 30, 70\\
    AGN fraction &0, 0.01, 0.1 - 0.9 (step 0.1), 0.99\\
    Extinction law of the polar dust & SMC \\
    E(B-V) for the extinction in the polar direction in magnitudes
    & 0, 0.2, 0.4\\
    \hline
     \multicolumn{2}{c}{X-ray : from AGN, galaxy \citep{stalevski2016dust}}\\
    Photon index $\Gamma$ of the AGN intrinsic X-ray spectrum & 1.8\\
   Power law slope connecting $L_\nu$ at rest frame $2500\AAA$ and 2 keV : $\alpha_\mathrm{{ox}}$ & $-1.9$, $-1.7$, $-1.5$, $-1.3$, $-1.1$\\
   Maximum allowed deviation of $\alpha_\mathrm{{ox}}$ from the empirical $\alpha_\mathrm{{ox}}-L_\nu$ 2500 \AA  & 0.2 \\
   \hline
    
  \end{tabular}
 \end{table*}

\begin{figure*}[htbp]
    
    \begin{tabular}{cc}
      \begin{minipage}[t]{0.42\hsize}
        \centering
        \includegraphics[keepaspectratio, scale=0.53]{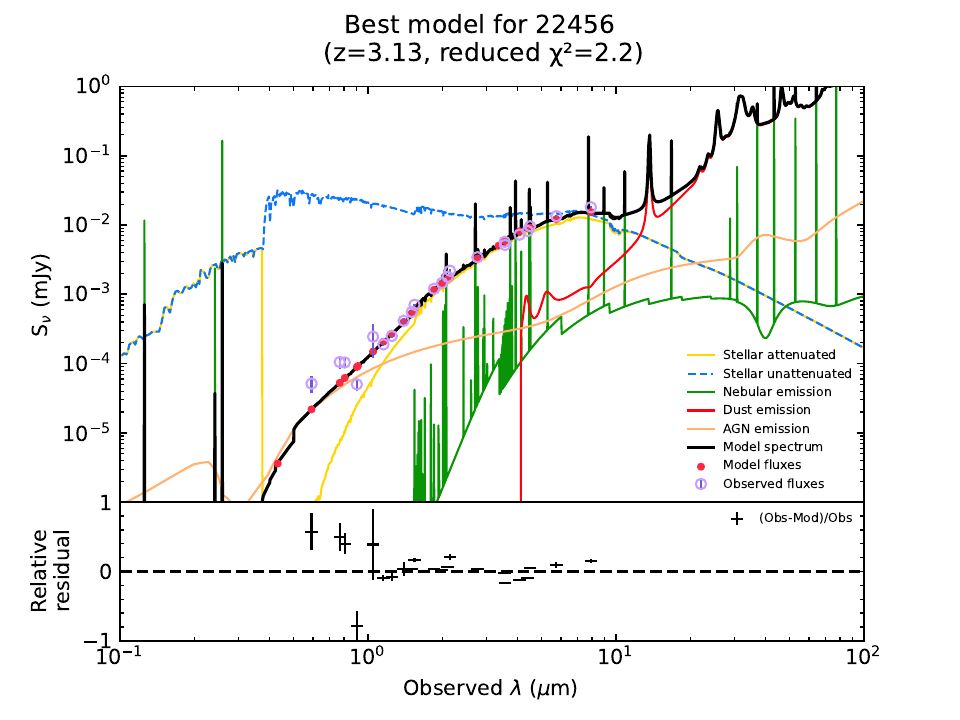}
      \end{minipage} &
      \begin{minipage}[t]{0.42\hsize}
        \centering
        \includegraphics[keepaspectratio, scale=0.53]{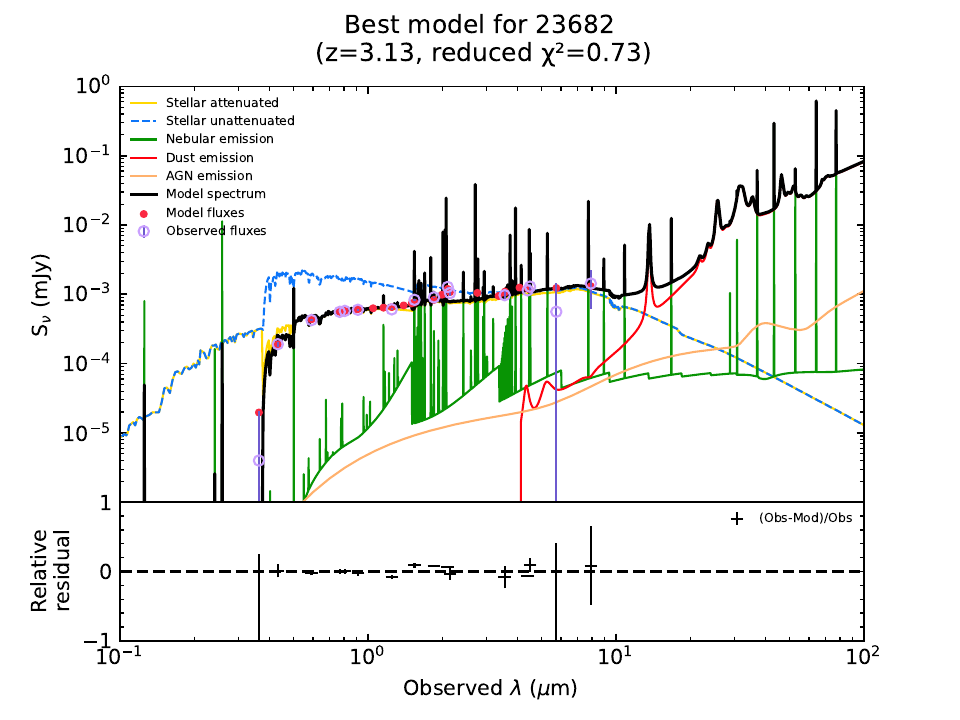}
      \end{minipage} \\
      \begin{minipage}[t]{0.42\hsize}
        \centering
        \includegraphics[keepaspectratio, scale=0.53]{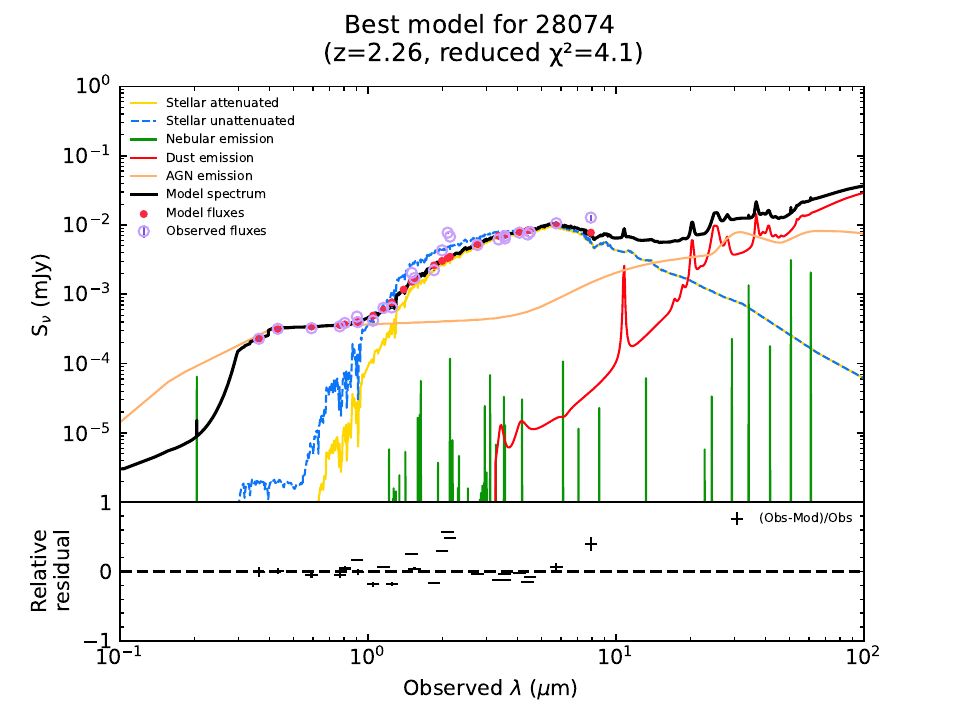}
      \end{minipage} &
      \begin{minipage}[t]{0.42\hsize}
        \centering
        \includegraphics[keepaspectratio, scale=0.53]{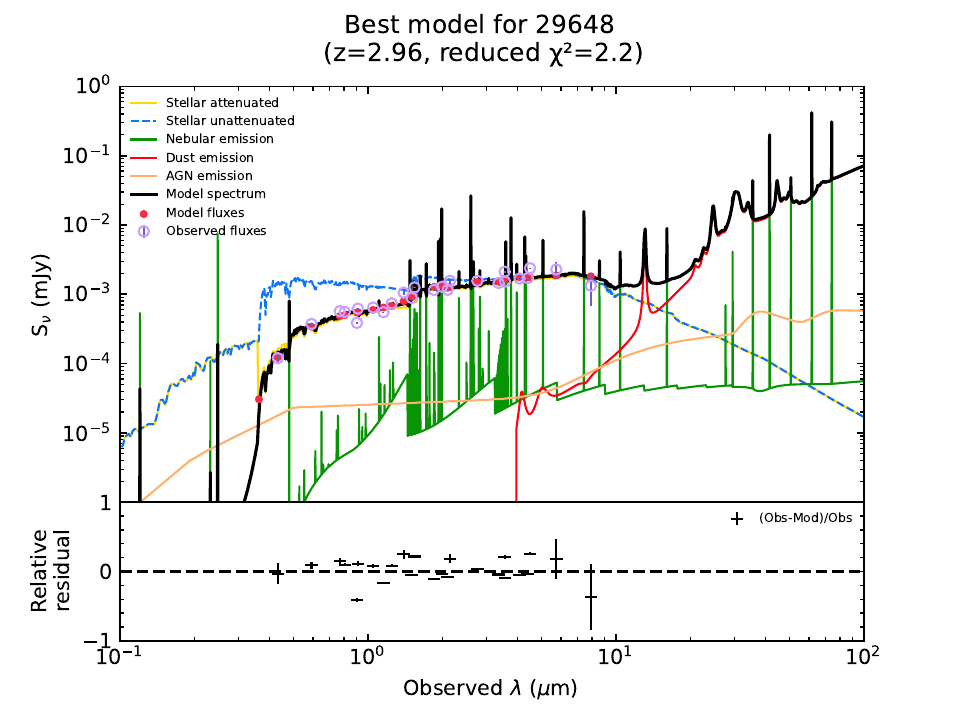}
      \end{minipage} \\
      \begin{minipage}[t]{0.42\hsize}
        \centering
        \includegraphics[keepaspectratio, scale=0.53]{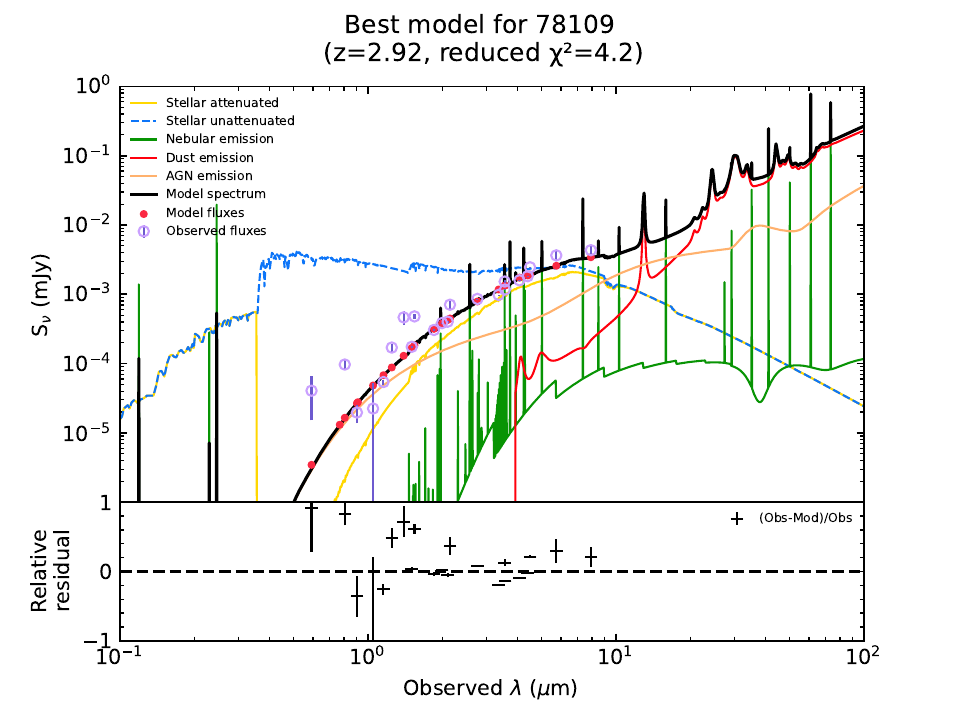}
      \end{minipage} &
       \begin{minipage}[t]{0.42\hsize}
        \centering
        \includegraphics[keepaspectratio, scale=0.53]{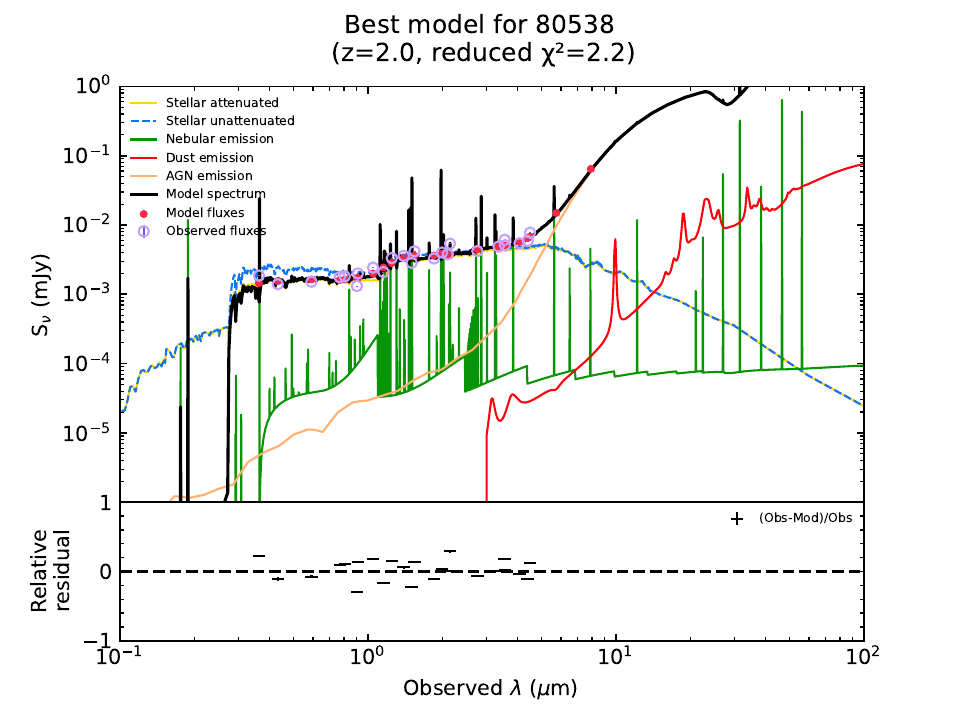}
      \end{minipage} \\
    \end{tabular}
    \caption{SED fitting results using CIGALE}\label{sed-plot1}
  \end{figure*}

 \begin{figure*}[htbp]
    
    \begin{tabular}{cc}
      \begin{minipage}[t]{0.42\hsize}
        \centering
        \includegraphics[keepaspectratio, scale=0.53]{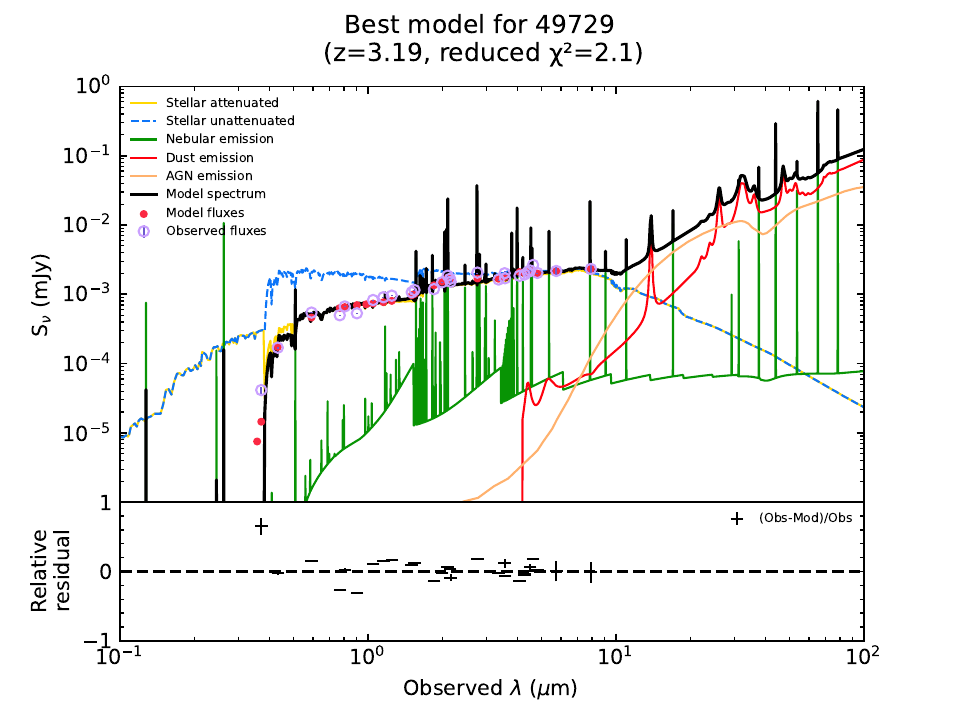}
      \end{minipage} &
      \begin{minipage}[t]{0.42\hsize}
        \centering
        \includegraphics[keepaspectratio, scale=0.53]{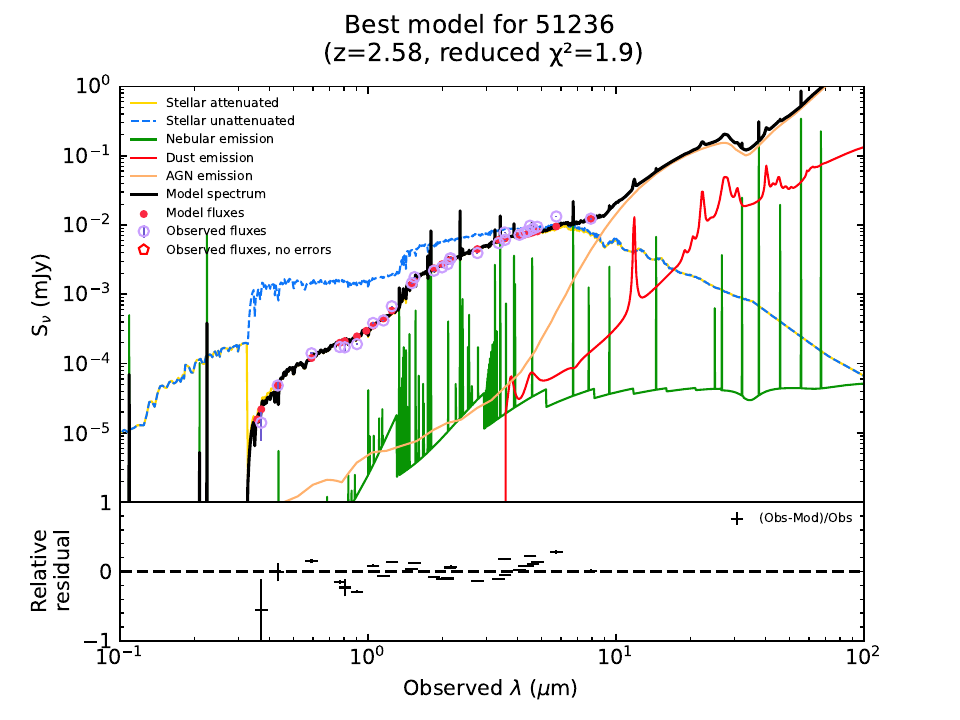}
      \end{minipage} \\

      \begin{minipage}[t]{0.42\hsize}
        \centering
        \includegraphics[keepaspectratio, scale=0.53]{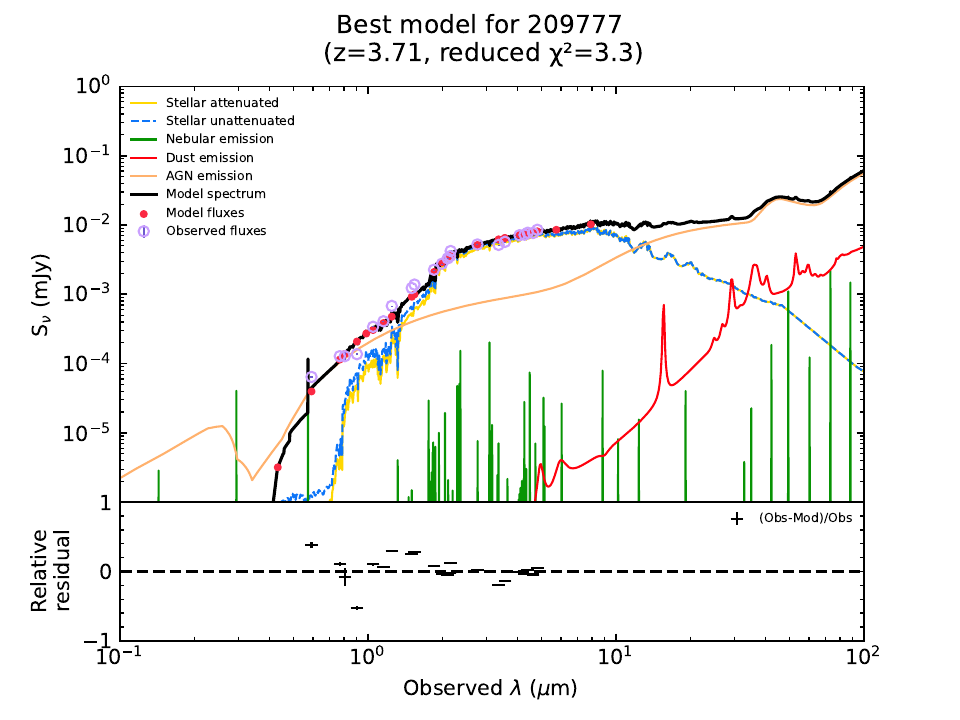}
      \end{minipage}
    \end{tabular}
    \caption{Continued}\label{sed-plot2}
  \end{figure*}

\begin{table*}[t]
 \centering
 \caption{Summary of Galfitting results in the F444W band}
  \label{table galfit AGNfrac}
   \begin{tabular}{cccccc}
    \hline \hline
    NIRSpec ID & Model & PSF mag & \sersic\ mag  & $\reff$ & $n$ \\
      &  & [mag] & [mag] & [pixel] &    \\
    \hline
    (1) & (2) & (3) & (4) & (5) & (6)  \\

\hline
22456 & 2\sersic & - & 22.12 $\pm$ 0.01& 3.26 $\pm$ 0.03 & 2.82 $\pm$ 0.07  \\
 &  & - & 22.61$\pm$ 0.02& 9.09 $\pm$ 0.07 & 0.26 $\pm$ 0.01   \\
23682 & \sersic & - & 24.23 $\pm$ 0.01& 5.37 $\pm$ 0.05 & 0.39 $\pm$ 0.02\\
28074 & PSF+\sersic & 24.11$\pm$ 0.04 & 21.95 $\pm$ 0.01& 1.61 $\pm$ 0.01 & 0.28 $\pm$ 0.03\\
29648 & 2\sersic & - & 23.89 $\pm$ 0.03& 6.11 $\pm$ 0.24 & 2.38 $\pm$ 0.09 \\
 &  & - & 23.92$\pm$ 0.02&17.98 $\pm$ 0.24 & 0.37 $\pm$ 0.02 \\
78109 & \sersic & - & 23.15$\pm$ 0.01 & 7.72 $\pm$ 0.08& 1.86$\pm$ 0.02\\
80538 & PSF+\sersic & 23.37 & 22.04$\pm$ 0.01 & 13.38 $\pm$ 0.11& 1.19$\pm$ 0.02 \\

\hline
49729 & PSF+\sersic & 25.97$\pm$ 0.05 & 23.25 $\pm$ 0.00& 1.78 $\pm$ 0.01 & 3.13 $\pm$ 0.05 \\
51236 & PSF+\sersic & 24.52$\pm$ 0.03 & 21.74 $\pm$ 0.00& 2.73 $\pm$ 0.01 & 1.89 $\pm$ 0.01\\
209777 & PSF+\sersic & 23.97$\pm$ 0.02 & 21.89 $\pm$ 0.00& 1.64 $\pm$ 0.01 & 2.89 $\pm$ 0.03\\
\hline

\end{tabular}
\begin{tablenotes}
\item[]Note: The errors shown here represent the fitting errors in Galfit. (1) NIRSpec ID (2) The best fitting model components (3)(4) Magnitude of PSF and \sersic component (5) Effective radius (6) \sersic index
\end{tablenotes}
\end{table*}

\begin{figure*}[htbp]
    \centering
    \begin{tabular}{ccc}
        \begin{minipage}[t]{0.3\hsize}
            \centering
            \includegraphics[keepaspectratio, scale=0.35]{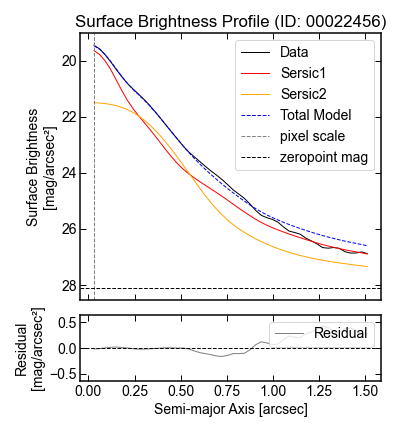}
        \end{minipage} &
        \begin{minipage}[t]{0.3\hsize}
            \centering
            \includegraphics[keepaspectratio, scale=0.35]{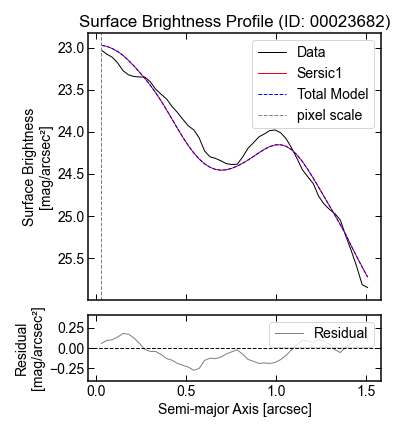}
        \end{minipage} &
        \begin{minipage}[t]{0.3\hsize}
            \centering
            \includegraphics[keepaspectratio, scale=0.35]{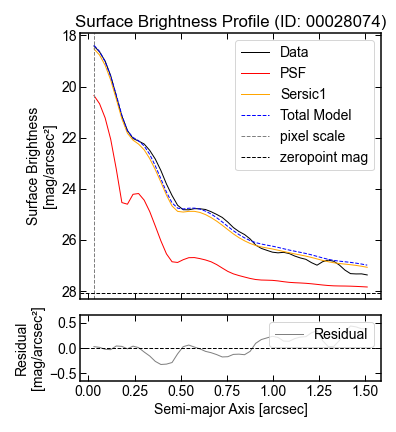}
        \end{minipage} \\
        \begin{minipage}[t]{0.3\hsize}
            \centering
            \includegraphics[keepaspectratio, scale=0.35]{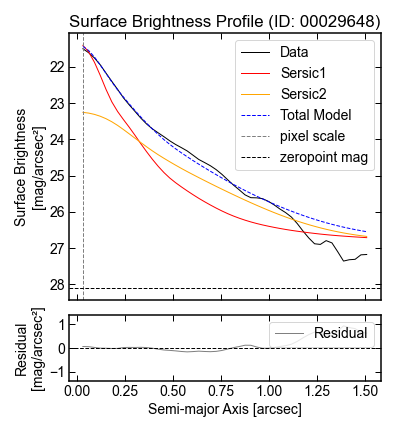}
        \end{minipage} &
        \begin{minipage}[t]{0.3\hsize}
            \centering
            \includegraphics[keepaspectratio, scale=0.35]{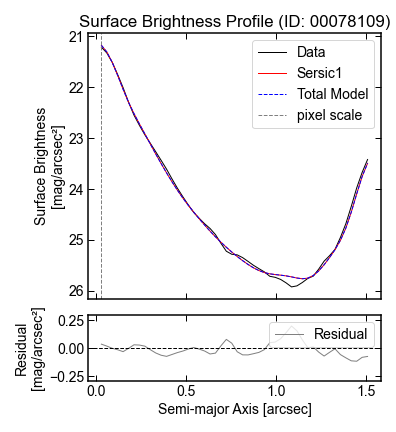}
        \end{minipage} &
        \begin{minipage}[t]{0.3\hsize}
            \centering
            \includegraphics[keepaspectratio, scale=0.35]{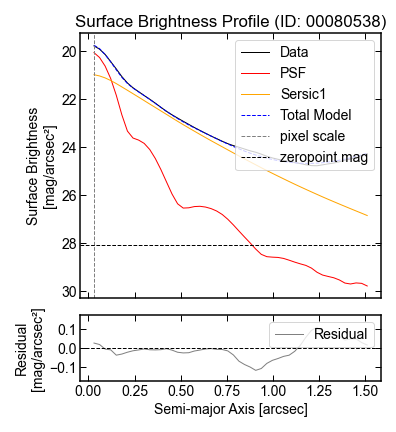}
        \end{minipage} \\
        \begin{minipage}[t]{0.3\hsize}
            \centering
            \includegraphics[keepaspectratio, scale=0.35]{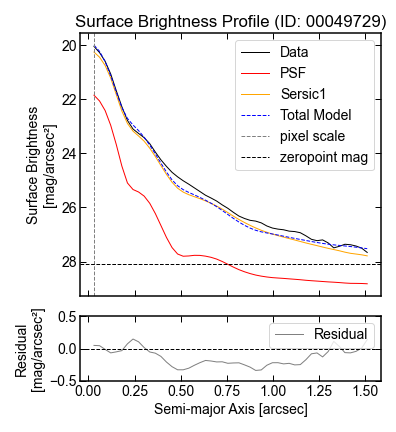}
        \end{minipage} &
        \begin{minipage}[t]{0.3\hsize}
            \centering
            \includegraphics[keepaspectratio, scale=0.35]{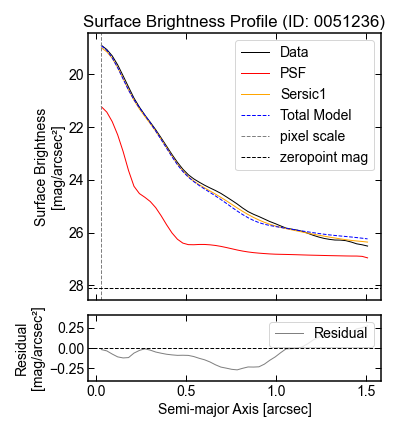}
        \end{minipage} &
        \begin{minipage}[t]{0.3\hsize}
            \centering
            \includegraphics[keepaspectratio, scale=0.35]{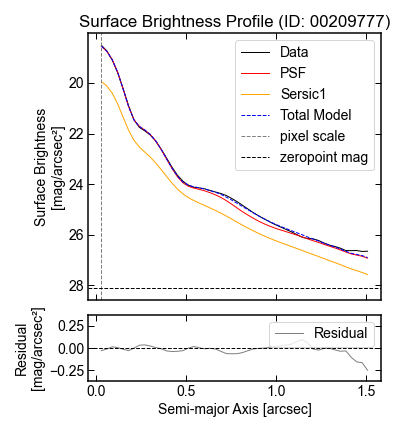}
        \end{minipage} \\
    \end{tabular}
    \caption{Radial surface brightness profiles of the target in the F444W band. The black solid line represents the observed data. The red and orange lines represent the PSF and Sérsic components, respectively. For galaxies modeled with two Sérsic components, the red and green solid lines show the first and second Sérsic components, respectively. The total model profile (sum of all fitted components) is shown as a blue dashed line. The vertical dashed gray line denotes the pixel scale, and the horizontal dashed black line indicates the zeropoint magnitude of the F444W band. The bottom panel shows the residuals between the observed profile and the total model (Data - Total). Each profile is measured along the semi-major axis using concentric elliptical annuli matched to the GALFIT best-fit geometry.}
    \label{fig-SB}
\end{figure*}

\end{document}